%% file: main.tex
\documentclass[a4paper,10pt,twocolumn]{article}

\usepackage{cite}
\usepackage{bmpsize}
\usepackage[colorlinks=true,urlcolor=black]{hyperref}
\def\BibTeX{{\rm B\kern-.05em{\sc i\kern-.025em b}\kern-.08em
    T\kern-.1667em\lower.7ex\hbox{E}\kern-.125emX}}

\usepackage{authblk}
\usepackage{abstract}
\usepackage{geometry}
\usepackage{amsmath,amssymb,amsfonts}
\usepackage{booktabs}
\usepackage{soul}
\usepackage{algorithm}
\usepackage{algorithmicx}
\usepackage[noend]{algpseudocode}
\usepackage{enumitem}
\usepackage{tikz}
\usepackage{graphicx}
\graphicspath{{./images/}}
\usepackage{float}

\usepackage{comment}

\usepackage{subcaption}

\usepackage{makecell}
\usepackage{multirow}
\usepackage{booktabs}

\usepackage{bm}

\usepackage{textcomp}

\usepackage[breakable]{tcolorbox}

\usepackage{xspace}

\newcommand{\safe}{SAFE\xspace}
\newcommand{\gemini}{Gemini\xspace}
\newcommand{\gmn}{GMN\xspace}
\newcommand{\catalog}{Catalog1\xspace}
\newcommand{\zeek}{ZEEK\xspace}
\newcommand{\binfinder}{BinFinder\xspace}
\newcommand{\jtrans}{jTrans\xspace}
\newcommand{\trex}{Trex\xspace}
\newcommand{\palmtree}{PalmTree\xspace}

\newcommand{\mFadv}{f_{adv}}
\newcommand{\fadv}{$\mFadv$\xspace}
\newcommand{\mPool}{P}
\newcommand{\pool}{$\mPool$\xspace}

\newcommand{\greedy}{greedy\xspace}

\newcommand{\mFq}{f_Q}
\newcommand{\fq}{$\mFq$\xspace}
\newcommand{\topk}{top-$K$\xspace}
\newcommand{\mV}{V}
\newcommand{\variants}{$\mV$\xspace}
\newcommand{\mFv}{f_v}
\newcommand{\fv}{$\mFv$\xspace}
\newcommand{\asr}{\textbf{ASR}\xspace}
\newcommand{\asri}{\textbf{ASR@\textit{i}}\xspace}
\newcommand{\asrone}{\textbf{ASR@\textit{1}}\xspace}
\newcommand{\asrtwo}{\textbf{ASR@\textit{2}}\xspace}
\newcommand{\asrthree}{\textbf{ASR@\textit{3}}\xspace}
\newcommand{\asrfour}{\textbf{ASR@\textit{4}}\xspace}
\newcommand{\wasr}{\textbf{wASR}\xspace}
\newcommand{\clean}{\textbf{INIT}\xspace}
\newcommand{\cleani}{\textbf{INIT@\textit{i}}\xspace}
\newcommand{\cleanone}{\textbf{INIT@\textit{1}}\xspace}

\newcommand{\cleanfour}{\textbf{INIT@\textit{4}}\xspace}
\newcommand{\modi}{\textbf{M-Instrs}}
\newcommand{\modn}{\textbf{M-Nodes}}

\newcommand{\recall}{\textbf{recall@\textit{10}}\xspace}
\newcommand{\recallK}{\textbf{recall@\textit{K}}\xspace}
\newcommand{\crecall}{\textbf{Recall pre attack}\xspace}
\newcommand{\arecall}{\textbf{Recall post attack}\xspace}
\newcommand{\trate}{\textbf{TSR}\xspace}
\newcommand{\vrate}{\textbf{VR}\xspace}

\definecolor{lightpurple}{RGB}{180,145,255}
\definecolor{mygreen}{RGB}{112,191,73}
\definecolor{Gainsboro}{rgb}{0.86, 0.86, 0.86}

\newtcolorbox{mybox}{
	arc=0pt,
	boxrule=0pt,
	colback=Gainsboro,
	width=\columnwidth,   % this option controls the width of the box
	colupper=black
}

\definecolor{babyblueeyes}{rgb}{0.63, 0.79, 0.95}

\newtcolorbox{mybox2}{
	arc=0pt,
	boxrule=0pt,
	breakable,
	colback=babyblueeyes,
	width=\columnwidth,   % this option controls the width of the box
	colupper=black
}

\title{\bfseries On the Lack of Robustness of Binary Function Similarity Systems}

\author[1]{Gianluca Capozzi}
\author[2]{Tong Tang}
\author[2]{Jie Wan}
\author[2,3]{Ziqi Yang}
\author[1]{Daniele Cono D'Elia}
\author[1]{Giuseppe Antonio Di Luna}
\author[4]{Lorenzo Cavallaro}
\author[1]{Leonardo Querzoni}

\affil[1]{Sapienza University of Rome, Italy\\ \texttt{\{capozzi,delia,diluna,querzoni\}@diag.uniroma1.it}}
\affil[2]{The State Key Laboratory of Blockchain and Data Security, Zhejiang University, China\\ \texttt{\{tong.tang, wanjie, yangziqi\}@zju.edu.cn}}
\affil[3]{Hangzhou High-Tech Zone (Binjiang) Institute of Blockchain and Data Security, China}
\affil[4]{University College London, UK\\ \texttt{l.cavallaro@ucl.ac.uk}}

\date{}

\begin{document}

\twocolumn[
  \maketitle
  \begin{center}
    \begin{minipage}{0.95\textwidth}
      \begin{abstract}
        \noindent
        Binary function similarity, which often relies on learning-based algorithms to identify what functions in a pool are most similar to a given query function, is a sought-after topic in different communities, including machine learning, software engineering, and security. Its importance stems from the impact it has in facilitating several crucial tasks, from reverse engineering and malware analysis to automated vulnerability detection. Whereas recent work cast light around performance on this long-studied problem, the research landscape remains largely lackluster in understanding the resiliency of the state-of-the-art machine learning models against adversarial attacks. As security requires to reason about adversaries, in this work we assess the robustness of such models through a simple yet effective black-box greedy attack, which modifies the topology and the content of the control flow of the attacked functions. We demonstrate that this attack is successful in compromising all the models, achieving average attack success rates of 57.06\% and 95.81\% depending on the problem settings (targeted and untargeted attacks). Our findings are insightful: top performance on clean data does not necessarily relate to top robustness properties, which explicitly highlights performance-robustness trade-offs one should consider when deploying such models, calling for further research.

        \vspace{1em}
      \end{abstract}
    \end{minipage}
  \end{center}
  \vspace{1.5em}
]

% Main content
\input{intro}

\input{model}
\input{solution}
\input{target}
\input{evaluation}
\input{related_work}

\section*{Acknowledgments}
This work was partially supported by the Italian MUR National Recovery and Resilience Plan funded by the European Union -- NextGenerationEU through projects SERICS (PE00000014) and Rome Technopole (ECS00000024).

\bibliographystyle{plain}
\bibliography{bibliography}

\appendix
\input{appendix}

\end{document}

%% file: intro.tex
% !TEX root =  main.tex

\section{Introduction}
\label{sec:introduction}

A fruitful and long-standing research trend involves applying Deep Neural Networks (DNNs) to solve binary analysis problems. These solutions typically provide end-to-end capabilities for handling complex tasks across entire binaries (prototypical examples include malware/benign classification solutions). More recently, the focus has narrowed to specific binary analysis challenges that could immediately assist a reverse engineer, such as decompiling binary functions~\cite{cao2022boosting}, identifying the signature and boundaries of a function~\cite{artuso2019nomine, 10.1145/3427228.3427265}, and detecting the toolchain used to generate a specific binary~\cite{massarelli2019investigating}.

\subsubsection*{The binary function similarity problem}
Among these tasks, one that has been predominately studied involves identifying when two binary functions are obtained from the same source code compiled with different compilers or optimization flags. This is known in the literature as the binary function similarity problem~\cite{alrabaee2015sigma, DBLP:conf/pldi/DavidPY16, dullien2005graph, khoo2013rendezvous}. This problem plays a key role in several security-sensitive scenarios~\cite{xu2017neural,massarelli2021function,marcelli2022machine}, and is especially effective in detecting previously analyzed functions using a reference database. This includes challenges such as identifying known library functions in statically linked stripped binaries, recognizing specific malware functionalities (e.g., by recognizing a particular crypto routine, or clustering malware into families and lineage trees), detecting known vulnerabilities in binaries and firmware, and identifying copyright infringement cases in compiled binaries.

All binary function similarity models take a pair of functions as input and output a similarity score, that ranges from a minimum to a maximum value. Even if the models are trained using the strict definition of similarity described above, it has been observed that high similarity scores are also given to functions derived from source codes that are different but semantically similar. This characteristic is indeed desirable as it can be used to cluster semantically similar functions.

The gold standard for testing binary function similarity solutions uses them in the \textbf{function search} problem~\cite{ding2019asm2vec, jTrans-ISSTA22, 10.1145/3579856.3582818}, where a query function \fq is used to order a pool of functions \pool according to their similarity score from \fq, where \pool can contain one or more %\LQ{why source? these are compiled functions...} 
functions \fv similar to \fq . The problem is correctly solved when \fv is among the \topk similar functions in the order induced by the similarity score.

Given its central importance, the binary function similarity problem has become a hot topic of research with various solutions, mainly based on DNNs, proposed in the last four years~\cite{xu2017neural, ding2019asm2vec, massarelli2021function, 10.1145/3579856.3582818, artuso2022binbert, shalev2018binary, DBLP:conf/ndss/DuanLWY20, 10.1145/3460120.3484587, DBLP:conf/icml/LiGDVK19, jTrans-ISSTA22, DBLP:journals/tse/PeiXYJR23}. To bring order to the plethora of proposed systems tested with varying performance measures, a 2022 paper by Marcelli et al.~\cite{marcelli2022machine} evaluated several solutions using a common dataset. This step represented the first attempt to systematize a still-growing and fascinating field (since 2022, other binary similarity models have been proposed~\cite{10.1145/3579856.3582818, jTrans-ISSTA22}).  % , 10.1145/3460120.3484587

\subsubsection*{The missing piece of the puzzle: robustness}
While~\cite{marcelli2022machine} systematically evaluates many different systems under several aspects, it never assesses the {\em robustness} of their underlying models.

A key weakness of machine learning solutions, especially those based on DNNs, is their performance when processing {\em adversarial examples}~\cite{yuan2019adversarial}. It is well-known that systems classifying media content (images, text, video, or audio) can be fooled by crafted examples obtained by modifying a benign one. Although the literature on adversarial examples is well-established for these models~\cite{DBLP:conf/pkdd/BiggioCMNSLGR13,goodfellow2014explaining,carlini2017towards,DBLP:conf/nips/0001QLSKM19,DBLP:conf/emnlp/JiaL17}, its investigation into systems that analyze binaries is still in the early stages, with the majority of works focused on fooling malware classifiers~\cite{pierazzi2020intriguing, lucas2021malware, demetrio2021functionality}.

%% ORIGINAL VERSION
%At the current state is unclear what would be the resiliency of the binary similarity models benchmarked in~\cite{marcelli2022machine} against adversarial example. An extensive evaluation of the robustness of binary similarity models against adversarial samples is missing and, in our opinion, is needed to complement the study of~\cite{marcelli2022machine}.

At the current state, it is unclear what would be the resiliency of the binary similarity models benchmarked in~\cite{marcelli2022machine} against adversarial examples. Intuitively, the ease of generating such examples for an adversary directly impacts the reliability of these systems. Hence, an extensive evaluation of their robustness is necessary to expose any inherent weaknesses undermining their practical value.

\subsubsection*{Our robustness evaluation}
In this paper, we aim to close this gap by investigating the robustness of binary function similarity models. We adopt a black-box approach, motivated by the objective of testing the models against the weakest possible adversary. In keeping with this spirit, we have decided to use a basic framework for our attack—a greedy approach—which we have extended with a few refinements: a black-box importance mechanism to decide which part of the function to modify, and an embedding-based mechanism for sequences of instructions to guide the content of certain transformations.

% The attacks that our adversary will mount can be \emph{targeted} or \emph{untargeted}.

In this paper, we consider an attacker aiming to compromise a binary function similarity system used for function search by reducing its ability to search for variants of a specific query function that they are altering. The adversary can execute \textit{targeted} and \textit{untargeted} attacks. In a targeted attack, given a query function \fq and a set \variants of functions that are semantically similar to each other, the attacker seeks to generate from \fq a semantically equivalent function \fadv maximizing its similarity with the functions in \variants. The untargeted attack is dual; here, given a query function \fq and a set \variants of functions semantically equivalent to \fq, the attacker seeks to generate from \fq a function \fadv semantically similar to \fq that minimizes the similarity with the functions in \variants.

%In a targeted attack, given a query function \fq, a pool of functions \pool, and a set $\mV \subset \mPool$ of semantically similar functions, the goal is to craft an adversarial function \fadv, semantically equivalent to \fq, that causes the BCSD system to rank the functions in \variants among the \topk most similar to \fadv. In an untargeted attack, the objective is to generate from \fq a function \fadv such that the functions in \variants, which are semantically equivalent to \fq, are excluded from the top similar results by the BCSD system.

%Specifically, our adversary can execute both \textit{targeted} and \textit{untargeted} attacks. Given a query function \fq, a pool \pool of functions, and a set $\mV \subset \mPool$ of semantically similar functions, the goal of the attacker in a targeted scenario is to create a function \fadv semantically equivalent to \fq and forces the BCSD system to rank the functions in \variants among the \topk functions in \pool most similar to \fadv. In an untargeted attack, the attacker is given a query function \fq, a pool \pool and a set $\mV \subset \mPool$ of functions semantically equivalent to \fq. Its goal is to create an adversarial version \fadv of \fq such that all the functions in \variants will not be ranker by the target BCSD model among its top similar functions.

We selected eight models, chosen from those used in~\cite{marcelli2022machine} and other more recent and promising ones, based on criteria of scalability and diversity. That is, the models must be scalable and thus usable in real-world settings, and they must cover a broad spectrum of potential characteristics of similarity models. These include, for example, manual and automatic features, different neural architectures (i.e., RNNs, feedforwards, and GNNs), models trained with and without execution information and with or without obfuscated samples. 
The diversity of models ensured during the selection process makes the evaluation in our paper generalizable, as the trends observed in our evaluation are likely to hold for other models that share the structure of some of those we tested.

The selected models have been tested against targeted and untargeted attacks using the black-box methodology described above, and their robustness has been evaluated according to the primary metric of the Attack Success Rate (\asr).

\subsection{Contributions}
In this paper, we assess the robustness of eight binary function similarity models---\gemini~\cite{xu2017neural}, \gmn~\cite{DBLP:conf/icml/LiGDVK19},  \zeek~\cite{shalev2018binary}, \binfinder~\cite{10.1145/3579856.3582818}, \safe~\cite{massarelli2021function}, \jtrans~\cite{jTrans-ISSTA22}, \trex~\cite{DBLP:journals/tse/PeiXYJR23}, and \palmtree~\cite{10.1145/3460120.3484587}--- using a simple black-box greedy attack. This attack leverages four semantics-preserving transformations that alter the topology and the content of the control flow graph of the attacked query function. 

This paper proposes the following contributions:
\begin{itemize}
    \item Robustness analysis. We assess the robustness of the examined models against targeted and untargeted attacks by using pools of various sizes and different values for $K$, which represent the number of functions returned by the model that are more similar to the query function \fq. We observed a significant difference in the Attack Success Rate (\textbf{ASR}) with targeted attacks being successful in about 57.06\% of cases, whereas untargeted attacks in 95.81\%.
    %\item Generalizability. Interestingly, the set of transformations utilized in our simple attack effectively addressed the key techniques and characteristics implemented by the attacked binary function similarity models. In the untargeted scenario, the ASR ranges from 98.30\% for the weaker model, \safe, to 92.99\% for the stronger model, \gemini. In contrast, the ASR radically decreases in the targeted scenario, with \binfinder showing the least robustness at 88.95\%, while \zeek shows the highest robustness with an ASR of 26.07\%.
    \item Transferability. We investigated whether an adversarial example crafted for one model could be used to attack another model. Our analysis shows that targeted attacks do not transfer well. On the other end, untargeted attacks generally transfer, with an average \textbf{ASR} of 62.11\%.
    \item Common model behaviors. We performed an in-depth analysis on the structure of the adversarial examples, to check whether they reveal useful insights about the attacked models.
    \item Our artifacts are available at: \url{https://github.com/Sap4Sec/BCSD_Robustness.git}
\end{itemize}

%% file: model.tex
\section{Threat Model}\label{ss:threat-model}
%In this section, we define our threat model and the problem of attacking binary similarity models.

This work focuses on assessing the robustness of binary function similarity systems at inference time (i.e., we do not investigate their robustness against poisoning attacks). To this end, we assume a black-box attacker~\cite{BIGGIO2018317, pierazzi2020intriguing}, with no access to the target model or training data. The attacker can perform an unlimited number of queries to observe the similarity value produced by the model. 

We emphasize that if the model does not provide the similarity score but only categorical outputs, the attack scenario shifts to a gray-box setting. However, the attacker's knowledge remains minimal, as they only need the similarity score to execute the attack effectively.

%\begin{figure}[h!]
%    \centering
%    \includegraphics[width=3.3in, trim = 0cm 1.0cm 0cm 0cm]{pool_for_query.pdf}
%    \caption{\textbf{(a)} \textbf{Targeted} attack representation. Initially, the target variants in $V = \{f_{v}^{1}, \cdots, f_{v}^{H}\}$ are not among the $K$ most similar functions of the pool \pool to \fq. After the attack, when using \fadv as a query, all the variants in $V$ enter the \topk ranking $P_K$. \textbf{(b)} \textbf{Untargeted} attack representation. Initially, the function \fq together with its variants $f_{v}^{1}, \cdots, f_{v}^{H}$ are among the $K$ most similar functions of the pool to \fq itself. At the end of the attack, when using its adversarial version \fadv as query, then \fq and its variants exit the top-$K$ ranking $P_K$.}
%    \label{img:TargetedAtt}
%    \vspace{-0.3cm}
%\end{figure} 

\begin{figure}[h]
	\centering
	\includegraphics[width=3.3in, trim = 0cm 0.2cm 0cm 0cm]{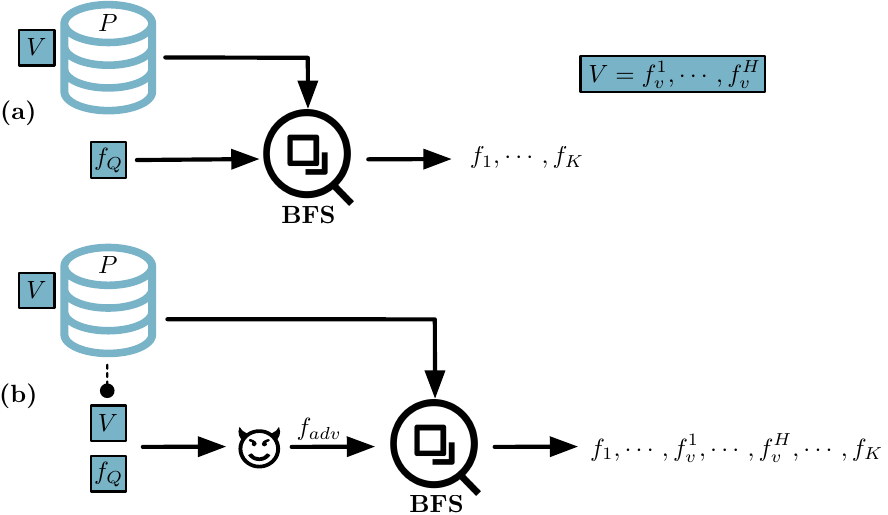}
	\caption{\textbf{Targeted} attack. \textbf{(a)} Initially, the target variants $V = \{f_{v}^{1}, \cdots, f_{v}^{H}\}$ are not among the \topk most similar functions to \fq in the pool \pool. \textbf{(b)} After the attack, using \fadv as query brings all variants in $V$ into the \topk.}
	\label{img:TargetedAtt}
\end{figure}

\begin{figure}[h]
	\centering
	\includegraphics[width=2.9in, trim = 0cm 0.2cm 0cm 0cm]{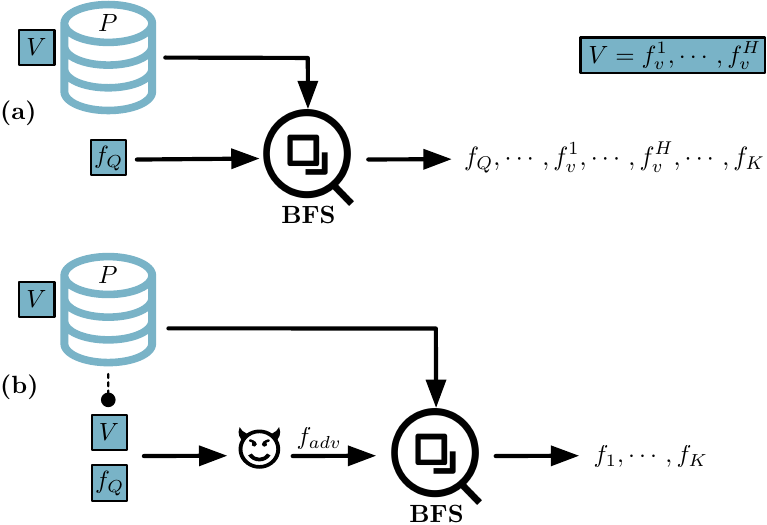}
	\caption{\textbf{Untargeted} attack. \textbf{(a)} Initially, \fq and its variants $V = \{f_{v}^{1}, \cdots, f_{v}^{H}\}$ are among the \topk most similar functions to \fq in the pool \pool. \textbf{(b)} After the attack, using \fadv as query removes \fq and its variants from the \topk.}
	\label{img:UntargetedAtt}
\end{figure}

\subsection{Targeted and Untargeted Attacks}\label{sec:problem-definition}
Let $sim$ be a similarity function that takes as input two functions and returns a real number, the {\em similarity score} between them. We define two binary functions as \textit{semantically equivalent} if they are two implementations of the same abstract functionality. A set of functions is a set of {\em variants} if all the functions are compiled from the same source code. 

An intriguing challenge within binary function similarity systems consists of the One-to-Many (OM) task~\cite{ding2019asm2vec, jTrans-ISSTA22, 10.1145/3579856.3582818} (or \textbf{function search}). In this task, given a pool of functions \pool, a query function \fq, and a number $K$ (where $K \le |P|$), we have to identify in \pool the $K$ functions that, according to the attacked model, are more similar to \fq. These $K$ functions are the \topk for the query \fq.

We consider two scenarios for an attacker interested in attacking the function search task: \textbf{targeted} and \textbf{untargeted} attacks.

In a \textbf{targeted attack}, the adversary is given a pool of functions, \pool, a set of target variants, $\mV \subset \mPool$, and a query function, \fq. The adversary has to find a function \fadv semantically equivalent to \fq. When \fadv is used as a query over \pool, the \topk must include $i$ target functions from \variants (with $i \leq K$).
 
A typical targeted attack occurs when the attacker wants to create an adversarial version \fadv of a specific malicious function \fq. This malicious function must resemble a certain benign function in \pool or one of its variants. Consequently, when the defender uses the binary function similarity system, a set of benign variants will be ranked among the \topk functions most similar to \fadv.
    
In a \textbf{untargeted attack}, the adversary is given a pool of functions, \pool, a query function, $\mFq \in \mPool$, and all its variants $V$ in \pool. The adversary has to find a function \fadv semantically equivalent to \fq such that, when \fadv is used as query over \pool, at least $i$ variants are not in the \topk (with $i \leq |V|$).
%the \topk must not contain the query function \fq and all of its variants.

A practical untargeted attack occurs when the attacker seeks to introduce a known vulnerable function, \fq, into a firmware. Their goal is to create a function, \fadv, that is semantically equivalent to \fq but as dissimilar as possible to all its variants, including \fq itself. As a result, when a binary function similarity system is used for vulnerability detection by the defender, none of the variants of \fq and \fq itself will be ranked among the \topk functions of \pool most similar to \fadv.

%let us consider the case where a BCSD system is used by a defender for identifying known vulnerabilities and an attacker that wants to introduce a vulnerable function \fq inside a firmware. In case of a targeted attack, the goal of the attacker is to generate a function \fadv semantically equivalent to \fq such that, when introduced in the firmware and presented to the defender, then it will not be detected by the BCSD model as similar to any of the variants of \fq that are in the \pool. In case of an untargeted attack, the goal of the attacker }

We present a visual representation of a targeted attack in Figure~\ref{img:TargetedAtt} and an untargeted attack in Figure~\ref{img:UntargetedAtt}.
 
%Another way to look at the attacks is by considering the success conditions with respect to the similarity score between \fadv and \ft. A targeted attack is successful when \( \text{sim}(\mFadv, \mFt) > \text{sim}(f_k, \mFt) \), where \( f_k \) is the least similar function to \ft among the top-\( k \). This formulation immediately implies that a targeted attack aims to create an adversarial sample that is more similar to \ft compared to the initial source \fs. The same reasoning applies to untargeted attacks, leading to the opposite conclusion: the adversarial sample has to be less similar to \ft compared to the initial source \fs. This alternative formulation will be the basis for designing the objective function that will guide our attack, details of which will be explained in Section~\ref{sec:objective}.

%% file: solution.tex
% !TEX root =  main.tex

\section{Attack Overview}\label{sec:blackbox}
In this section, we present the black-box procedure we utilize to assess the robustness of the models under consideration. We first describe the objective function the attacker seeks to solve and the optimization strategy adopted for generating adversarial examples. Then, we introduce the semantics-preserving techniques for manipulating binary functions we embody in our attack.

\subsection{Multi-Objective Optimization}\label{sec:objective}
In the following, we refer to the targeted attack case. The same rationale, with the necessary minor adjustments, holds for the untargeted case.

In the context of a targeted attack, given a query function \fq, and set $\mV \subset \mPool$ of target variants, the goal of the attacker consists of generating an adversarial example \fadv starting from \fq that maximizes the similarity between \fadv and all the target variants $\mFv \in \mV$. This translates into the following multi-objective function on an undefined number of variables (one for each variant):

\begin{equation}
    \max_{\mFadv} (sim(\mFadv, \mFv^1), \ldots, sim(\mFadv, \mFv^H))
\end{equation}

We solve this multi-objective problem by reducing it to the following max-min problem that takes into account also the perturbation size:

\begin{equation}\label{eq:obj}
    \max_{\mFadv} \; \min_{\mFv} \; sim(\mFadv, \mFv) - \lambda  \cdot \frac{|len(\mFq) - len(\mFadv)|}{len(\mFq)}
\end{equation}

\noindent
where:
\begin{itemize}
    \item $len(\cdot)$ takes a binary function as input and returns its length in terms of the number of instructions;
    \item $\lambda$ determines how much the size of the perturbation should penalize the produced adversarial example. In the following, we refer to $\lambda$ as the \textbf{penalty factor}. 
\end{itemize}
Informally, with this max-min problem, we maximize the minimum similarity between \fadv and the variants in $V$. 

However, we highlight that differently from the computer vision scenario~\cite{goodfellow2014explaining}, the perturbation size does not present a significant concern within our threat model. Indeed, as elucidated in~\cite{pierazzi2020intriguing}, an adversarial binary code must exhibit both validity and realism. Consequently, our set of transformations must alter \fadv so that it will still look plausible when manually analyzed, which doesn't imply minimizing the modification size.

\begin{figure*}
    \centering
    \includegraphics[width=0.90\linewidth]{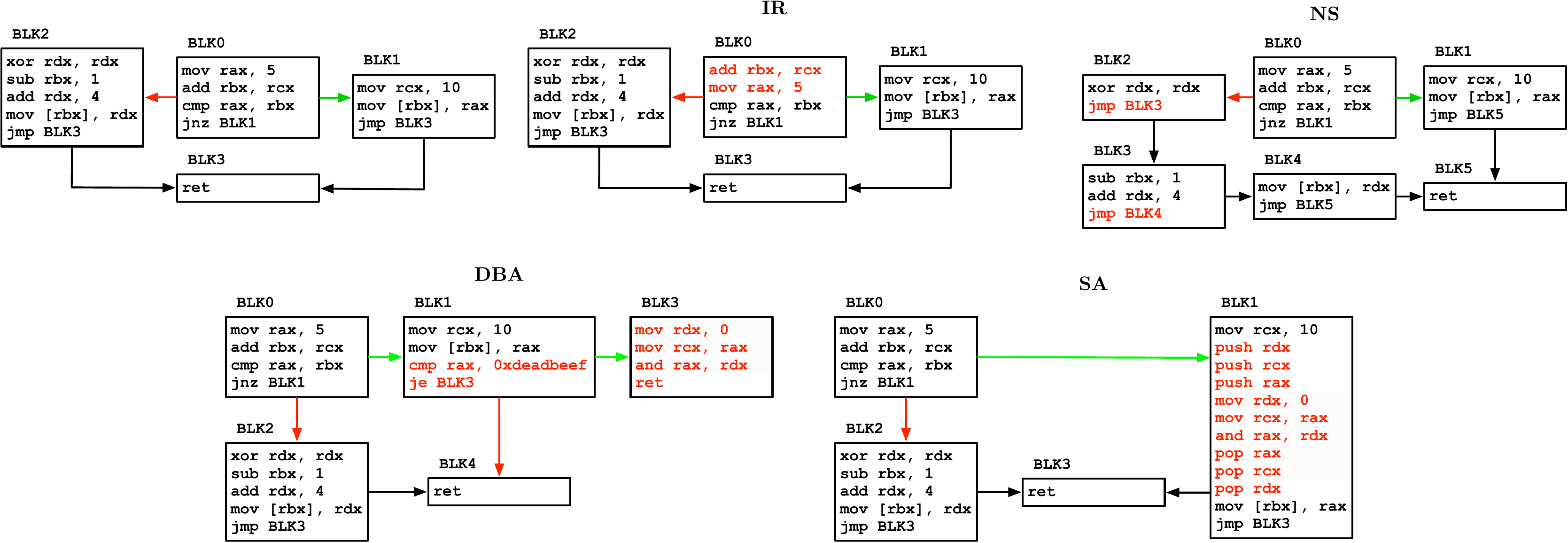}
    \caption{Semantics-preserving transformations embodied in the attack.}
    \label{fig:transformations}
\end{figure*}

\subsubsection{Greedy Optimizer}\label{sec:greedy}
To solve the max-min problem, we use an $\varepsilon$-greedy gradient-free optimizer to iteratively modify \fq toward the desired similarity outcome. That is, maximize the similarity between the adversarial example \fadv and the least similar variant in \pool.

% maximize the similarity between the adversarial example \fadv and the least similar variant in $V$.

% We solve our max-min problem leveraging an $\varepsilon$-greedy gradient-free optimizer to iteratively modify the original sample \fq toward the desired similarity outcome. That is, maximize the similarity between the adversarial example \fadv and the least similar variant in $V$.

At each iteration, starting from the output $\mFadv'$ of the previous iteration (or the original function for the first iteration), we generate multiple candidate adversarial examples. This is done by applying \textit{semantics-preserving} transformations from a predefined set $TR$ to specific locations within $\mFadv'$ identified in the set $POS$. Specifically, we create a fixed number of candidates by considering all possible pairs $\langle tr, pos \rangle$, where $tr \in TR$ and $pos \in POS$. Notably, a single transformation can generate multiple candidates when applied to a given location. The next section details the specific semantics-preserving transformations used in this process.

For each candidate, we compute the objective function in Equation~\ref{eq:obj}; then, we select the adversarial example \fadv for the next iteration using an $\varepsilon$-greedy procedure over the generated candidates. Specifically, with probability $1 - \varepsilon$, we select the candidate that maximizes the objective function, and with probability $\varepsilon$, we choose a suboptimal one. At the end of the iterative procedure, we select as the final \fadv the one that produced the highest value for the objective function among all \fadv generated at the end of each iteration. This is then used as a query over \pool.

%the optimization strategy outlined in Figure~\ref{img:BBwork}. This strategy consists of an $\varepsilon$-\greedy procedure that iteratively modifies the original sample \fq by applying \textit{semantics-preserving} transformations. 

% At each iteration, we generate several adversarial examples and select, with probability $1-\varepsilon$, the one that maximizes the similarity between \fadv and its least similar function in \variants. With probability $\varepsilon$, we choose a suboptimal example. This approach follows the standard $\varepsilon$-\greedy procedure, aimed at increasing the probability of escaping local optima.

% Each new example is generated by applying a semantics-preserving transformation to the adversarial example $\mFadv'$ selected in the previous iteration (with \fq being the initial $\mFadv'$). 

A certain semantics-preserving transformation can only be applied to a specific set of positions inside the original function we want to mutate (for example, we cannot swap instructions having a dependency). However, we have to further restrict this set to keep our attack computationally feasible. We do this by identifying the positions where transformations are more likely to greatly impact the similarity function. Here, each position is an assembly instruction within the function being modified. Specifically, the importance of instruction $i \in \mFadv'$ is the absolute variation of the similarity value of $\mFadv'$ and its least similar function $f_v \in V$ measured when removing $i$:

\begin{equation}
    IM_{i} = | sim(\mFadv', f_v) - sim(\mFadv' \setminus i, f_v)|,
\end{equation}

\noindent
where $\mFadv' \setminus i$ is the function $\mFadv'$ without instruction $i$.

%The importance of an instruction $i_t \in \mFadv'$ is the absolute variation of the similarity value of  $\mFadv'$ and its least similar function in \variants measured when removing $i_t$. 
After computing the importance score for each instruction $i \in \mFadv'$, the ones with the highest value will be candidates for applying \textit{semantics-preserving} perturbations.

\subsection{Semantics-Preserving Transformations}\label{sec:semPres}
As noted in~\cite{pierazzi2020intriguing}, an adversarial function \fadv, generated from a binary function \fq must satisfy \textbf{problem-space constraints}, including preserving its semantics.

% As described in~\cite{pierazzi2020intriguing}, given a binary function \fq, its adversarial version \fadv should adhere to \textbf{problem-space constraints}, among which there is the preservation of its semantics.

Figure~\ref{fig:transformations} illustrates the semantics-preserving transformation techniques we embodied in our attack strategy. Some of these were initially discussed in~\cite{capozzi2023adversarial}, together with a categorization based on whether they modify or not the control-flow graph (CFG). Among these transformations, we embed in our attack strategy: Instruction Reordering \textbf{(IR)}, Node Split \textbf{(NS)}, and Dead Branch Addition \textbf{(DBA)}. Specifically, \textbf{IR} reorders consecutive data-independent instructions within the function, ensuring that the swap preserves the semantics. \textbf{NS} splits an existing CFG node into three separate nodes using unconditional jumps, without altering the semantics of the function. Finally, \textbf{DBA} introduces dead code--- specifically, a strand (i.e., a sequence of data-dependent assembly instructions~\cite{artuso2022binbert})--- into a new basic block that is guarded by an always-false branch.

% In particular, \textbf{IR} swaps semantically independent instructions. The \textbf{NS} transformation splits an existing CFG node into three different nodes using unconditional jumps. Finally, \textbf{DBA} introduces a strand, i.e. a sequence of data-dependent assembly instructions~\cite{artuso2022binbert}, in a new node of the CFG. The original semantics is preserved by using an always-false condition in the conditional \textit{jump} that points to the inserted node.

Additionally, we introduce a new transformation called Strand Addition \textbf{(SA)}, which inserts a strand into an existing CFG node. The strands under consideration contain neither control flow nor memory-modifying instructions. Furthermore, using liveness analysis, we identify all registers and flags used within the strand, saving their contents at the beginning and restoring them at the end, to ensure that the function's semantics remain unaltered.

% we guarantee the preservation of the semantics by saving the content of all the registers used by the strand at the start and restoring it at the end.

We select the strand to add to the function by enumerating a set of candidates sampled from a large dataset of available strands. Rather than defining a fixed set of strands, we rely on an embedding space to establish a set of candidates that is dynamically updated at the end of each iteration within the optimization procedure. We initialize this set using strands uniformly sampled from our large dataset. At the end of each optimizer iteration, we update part of the set with strands selected from the closest neighbors (using the embedding space) of those representing the top \greedy actions from the previous iteration (specifically from the \textbf{SA} category), while the remaining portion is filled with new random strands.

We define the embedding space using a strand embedding model to transform strands into vectors. These vectors are grouped according to the semantics of the strands, allowing us to establish a notion of proximity between them. We choose BinBert~\cite{artuso2022binbert} as the model for generating strand embeddings because it has been specifically fine-tuned for strands similarity detection.

We provide a detailed pseudocode of our attack strategy in Appendix~\ref{sec:app_C}.

%% file: target.tex
% !TEX root = main.tex

\section{Target Systems}\label{sec:targetSystems}

In this section, we illustrate the binary function similarity models we attacked. Based on~\cite{marcelli2022machine}, we select the different solutions according to the following \textbf{selection criteria}:
\begin{itemize}
    \item \textbf{Scalability}. To demonstrate the potential vulnerabilities of models in a real-world scenario, we evaluate approaches that are not slow at inference time.
    \item \textbf{Diversity of proposed approach}. Binary function similarity solutions span multiple research communities (i.e., system security, programming languages, machine learning) and multiple approaches. Moreover, within each community, diverse techniques have emerged; for instance, the machine learning community explores various architectures like GNNs, RNNs, and Transformers. Therefore, we select a range of models that not only cover different architectural structures but have also been proposed by different research communities.
    %\item \textbf{Latest trends}. Most of the latest advancements in the state-of-the-art consider deep learning-based models for addressing the problem of binary similarity, which outperform traditional solutions on clean data. Thus, while we consider various approaches, we prioritize deep learning-based solutions.
\end{itemize}

\noindent
We now briefly describe the targeted models.

%\subsection{Bytes fuzzy hashing: Catalog1}
%\noindent
%\textbf{Catalog1} \cite{catalog} is a fuzzy hashing-based approach that directly uses the raw binary information. It collects the consecutive 4 bytes from one function and subsequently employs MinHash~\cite{broder1997resemblance} to encode the byte set into a constant-sized signature. Catalog1 algorithm assumes that the bytes set encapsulates enough information from the function. Ultimately, Jaccard similarity is used to compare the similarity between the two sets. This method is more effective for identifying the similar segments of the two functions in most practical cases.

\subsection{Graph Neural Network (GNN): \gemini and \gmn}

\noindent
\textbf{\gemini} (2017)~\cite{xu2017neural} is built on a graph neural network derived from the Structure2Vec~\cite{DBLP:conf/icml/DaiDS16} model and trained using a Siamese architecture. It converts an ACFG (a control flow graph with manual block-level features) into an embedding vector, which is generated by aggregating the embedding vectors of the individual nodes within the ACFG. Given two binary functions, their similarity is determined by the cosine similarity of their ACFG embedding vectors. %, which are learned using a Siamese architecture.

\noindent
\textbf{Graph Matching Network} (\gmn) (2019)~\cite{DBLP:conf/icml/LiGDVK19} consists of a graph neural network that calculates the similarity between two functions by representing them through their CFGs. Unlike standard GNN solutions (i.e., \gemini~\cite{xu2017neural}), where the embedding vector for a node captures properties from its neighborhood only, \gmn also searches for possible matches across the two input graphs. In particular, \gmn computes the distance between the inputs by trying to match their nodes.

\subsection{Intermediate Representation (IR) and Neural Network (NN): Zeek}
\noindent
\textbf{Zeek}~\cite{shalev2018binary} uses an intermediate representation to capture the semantics of the input functions. Specifically, starting from the CFG of a function, it first decomposes each basic block into strands and converts the assembly code in each strand to a normalized intermediate representation (i.e., VEX-IR). Subsequently, it generates a hash value for each block, which is then indexed in a vector to represent the function. Finally, it uses two fully connected hidden layers to detect the similarity between vectors derived from semantically equivalent code sections.

\subsection{Fully Connected Neural Network: \binfinder}

\noindent
\textbf{\binfinder} (2023)~\cite{10.1145/3579856.3582818} employs a fully connected neural network trained using a Siamese architecture to generate function embeddings, which are then employed for similarity calculation. It represents the input functions through a set of static features that are claimed to be robust to code obfuscation, compiler optimization, and cross-compilation processes. These features are first processed using different tokenizers; then, they are transformed into one-hot vectors.

%\subsection*{Assembly Code Embedding and PV-DM: Asm2Vec}
%\textbf{Asm2Vec}~\cite{ding2019asm2vec} is a function embedding model based on the PV-DM model~\cite{le2014distributed} for learning documents representation. Asm2Vec first extracts a series of execution traces from the CFG using random walks; then, leveraging the PV-DM model, it computes an embedding for each sequence while concurrently producing vector representations of single assembly instructions. 
% model the CFG as multiple sequences, where each sequence corresponds to a possible execution trace (which can be modeled as a sequence of assembly instructions).
% each operand and operation of an asm instruction is treated as a token.
% a PV-DM-based model is used to learn relationships among the instructions of the trace
% use edge sampling and random walks to generate instruction sequences

\subsection{Recurrent Neural Network (RNN): \safe}

\noindent
\textbf{\safe} (2022)~\cite{massarelli2021function} is a recurrent neural network-based model trained using a Siamese architecture, that converts the linear disassembly of the input functions into embedding vectors. It first computes an embedding for each assembly instruction using a model derived from the word2vec~\cite{DBLP:conf/nips/MikolovSCCD13} word embedding model. Then, using a self-attentive component, it aggregates these vectors into a final function embedding vector. Similar to \gemini~\cite{xu2017neural}, the similarity between two functions consists of the cosine distance between their corresponding embeddings.

\subsection{Transformer: \jtrans, \trex, \palmtree}

\noindent
\textbf{\jtrans} (2022)~\cite{jTrans-ISSTA22} is a BERT-based model~\cite{devlin2018bert} that combines instruction semantics with CFG information to infer the binary code representation. When considering an assembly instruction, \jtrans treats each mnemonic and operands as tokens and computes an embedding for each of them. However, to make \jtrans better in capturing the control flow execution, \textit{jump} instructions are modeled differently from other ISA instructions. The last layer of the model generates the function's embedding, which can then be used for similarity calculation. \jtrans has been pre-trained on two tasks (i.e., Masked Language Model and Jump Target Prediction) and then fine-tuned for Binary Similarity Detection.

\noindent
\textbf{\trex} (2023)~\cite{DBLP:journals/tse/PeiXYJR23} is a transfer-learning-based framework that adopts a hierarchical Transformer~\cite{DBLP:conf/sigsoft/PeiGBCYWUYRJ21} for learning functions' semantics via regular and forced execution traces. The model is first pre-trained using a Masked Language Modeling-like task. In particular, given a function's trace (which comprises both instructions and values), some parts are randomly masked to be predicted during model training using the surrounding context. Once pre-trained, the model is then fine-tuned for function similarity detection, using functions' static code (instead of traces) as input to the model. Specifically, the function's embedding is the output of a two-layer MLP. This MLP takes as input the mean pooling of the embeddings produced by the last self-attention layer of the fine-tuned model.

\noindent
\textbf{\palmtree} (2021)~\cite{10.1145/3460120.3484587} is a BERT-based~\cite{devlin2018bert} pre-trained assembly language model for generating general-purpose instruction embeddings. Here, assembly instructions are treated as separate sentences composed by basic tokens. In particular, each operand is decomposed into basic elements, normalizing strings and addresses with special tokens to avoid the OOV (Out-Of-Vocabulary) problem. The model is pre-trained considering three tasks: Masked Language Model, Context Window Prediction, and Def-Use Prediction. Since \palmtree is an instruction embeddings model, to evaluate its performance on the function similarity task, we follow the strategy used in \cite{artuso2022binbert}, where an LSTM aggregates the instruction embeddings produced by \palmtree into a function embedding.

%% file: evaluation.tex
% !TEX root = main.tex

\section{Datasets and Implementation}
This section describes the datasets we use to evaluate our approaches and implementation details.

\subsection{Dataset}\label{sec:dataset}
We test our approach by considering a codebase of binary functions extracted from 8 open-source projects written in C language: binutils, curl, openssl, sqlite, gsl, libconfig, ffmpeg, and postgresql. We compile the programs for the \texttt{amd64} architecture, considering two compilers (i.e., \texttt{gcc-9.4.0} and \texttt{clang-12}) and two optimization levels (namely, \texttt{O0} and \texttt{O3}), obtaining 4 different combinations. The collection we obtained reflects real-world software, including binaries utilized in the evaluation or training of the binary function similarity systems examined in this work. 

We extract the binary functions using \texttt{Radare2}~\cite{radare2} disassembler, excluding those that contain fewer than 6 assembly instructions or 2 CFG nodes, obtaining 127,534 functions. Our final codebase consists only of the functions for which there are exactly four variants, one for each combination of compiler and optimization level.

Once obtained the codebase, we create pools of different sizes. A pool \pool of size $S$ is composed of $S/4$ distinct functions uniformly sampled from the codebase together with their variants. Our final targeted dataset comprises 1,000 samples, each representing a query on a certain pool on which the attack has to be carried out. Specifically, each targeted sample contains: the pool \pool; a set $\mV \subset \mPool$ of target variants; a query function \fq (the one that will be modified by the attacker) extracted randomly among our codebase of functions. Note that $\mFq \not\in \mV$.

For the untargeted case, we create a similar dataset of 1,000 samples, each of which contains: the pool \pool; a set $\mV \subset \mPool$ of target variants;
 a query function $\mFq \in \mV$ (the one that will be modified by the attacker).

\subsection{Implementation Details}\label{sec:implementation}

We implement our attack as a two-phase procedure; the first one consists of disassembling the input functions and building a high-level representation using their CFGs; the latter consists of applying the transformations to the aforementioned representations and then feeding them in input to the target model. Therefore, we remark that our attack is not done using binary rewriting techniques but it is performed on this high-level representation. However, we want to stress that all our transformations are semantics-preserving \textit{by design}, as detailed in Section~\ref{sec:semPres}; furthermore, all these transformations can be effectively applied as source code modifications by first translating the C code of the function into assembly code (by compiling the \texttt{.c} file into a \texttt{.s} file) and then by directly modifying the obtained assembly code. In this way, the injection process will not rely on binary rewriting techniques, which are more prone to alter the semantics due to relocation issues. 

We built our CFG extraction module upon the \texttt{Radare2}\footnote{\url{https://github.com/radareorg/radare2}} and \texttt{angr}~\cite{DBLP:conf/sp/Shoshitaishvili16} disassemblers. 

Our testing pipeline has been coded in Python. We consider the official implementation and settings for each target model except for \gemini, \gmn, and \zeek for which we use alternative implementations.\footnote{Due to unavailability of original implementations or incompatibility with our pipeline.}

\section{Experimental Results} \label{sec:exp}
In this section, we provide the results of our experimental evaluation. We first define the performance metrics we use, then we describe our choice of the attack hyperparameters and finally, we evaluate the resiliency of binary similarity models by investigating the following research questions:

\begin{mybox}
    \textbf{RQ1}: \textit{Do the models exhibit greater robustness against targeted or untargeted attacks?}
    
    %\textit{Is it possible to deduce aspects of the model through a qualitative analysis of the generated adversarial example?}
    
    \textbf{RQ2}: \textit{Are the considered transformations able to generalize across various categories of target models? Do adversarial examples transfer across models?}

    \textbf{RQ3} \textit{Is it possible to deduce aspects of the model through the distribution of applied transformations?}

\end{mybox}

\begin{table*}
\centering
%\footnotesize
\caption{Untargeted attack at $K=10$ when considering pools with size 128 and 512 and setting $\lambda = 0$. In column AVG we report the average of the measures across all models.}
\resizebox{\textwidth}{!}{%
    \begin{tabular}{|c|c|rr|rr|rr|rr|rr|rr|rr|rr|rr|} 
        \cmidrule[\heavyrulewidth]{3-18}
        \multicolumn{1}{l}{\multirow{3}{*}{}} & \multicolumn{1}{l|}{\multirow{2}{*}{}} & \multicolumn{16}{c|}{\textbf{Models}} & \multicolumn{2}{c}{\multirow{1}{*}{}} \\
        \cline{3-20}
        \multicolumn{1}{l}{} & \multicolumn{1}{l|}{} & \multicolumn{2}{c|}{\gemini} & \multicolumn{2}{c|}{\gmn} & \multicolumn{2}{c|}{\zeek} & \multicolumn{2}{c|}{\binfinder} & \multicolumn{2}{c|}{\safe} & \multicolumn{2}{c|}{\jtrans} & \multicolumn{2}{c|}{\trex} & \multicolumn{2}{c|}{\palmtree} & \multicolumn{2}{c|}{AVG} \\
        \cline{2-20}
        \multicolumn{1}{l|}{} & \bm{$|\mPool|$} & \textbf{128} & \textbf{512} & \textbf{128} & \textbf{512} & \textbf{128} & \textbf{512} & \textbf{128} & \textbf{512} & \textbf{128} & \textbf{512} & \textbf{128} & \textbf{512} & \textbf{128} & \textbf{512} & \textbf{128} & \textbf{512} & \textbf{128} & \textbf{512} \\
        \toprule
        \toprule
        & \crecall & 0.71 & 0.56 & 0.94 & 0.86 & 0.73 & 0.51 & 0.91 & 0.83 & 0.94 & 0.84 & 0.81 & 0.67 & 0.95 & 0.89 & 0.90 & 0.79 & 0.86 & 0.74 \\
        & \arecall & 0.07 & 0.04 & 0.03 & 0.01 & 0.05 & 0.02 & 0.06 & 0.04 & 0.02 & 0 & 0.04 & 0.01 & 0.04 & 0.02 & 0.03 & 0.01 & 0.04 & 0.02 \\
        & \wasr & 92.99 & 96.27 & 97.30 & 99.0 & 95.23 & 98.12 & 93.80 & 96.33 & 98.30 & 99.58 & 96.0 & 99.15 & 96.17 & 98.43 & 96.69 & 99.37 & 95.81 & 98.28 \\
        \toprule
        \toprule
        & \clean & 63.63 & 82.87 & 16.10 & 36.40 & 61.76 & 84.70 & 24.40 & 39.40 & 19.40 & 42.60 & 51.10 & 73.20 & 12.82 & 29.22 & 27.66 & 53.15 & 34.61 & 55.19 \\
        & \asr & 96.69 & 98.10 & 99.30 & 99.70 & 98.06 & 99.09 & 95.80 & 97.80 & 99.20 & 100.0 & 99.40 & 100.0 & 97.61 & 99.30 & 98.0 & 99.90 & 98.01 & 99.23 \\
        & \modi & 235.07 & 234.18 & 214.44 & 215.0 & 201.77 & 202.11 & 226.76 & 227.07 & 237.61 & 236.85 & 133.19 & 133.06 & 195.51 & 193.39 & 524.32 & 522.24 & 246.08 & 245.49 \\
        \multirow{-4}{*}{\textsf{\textbf{@1}}} & \modn & 26.98 & 26.95 & 16.90 & 16.94 & 25.68 & 25.83 & 41.73 & 41.80 & 13.68 & 13.66 & 16.74 & 16.73 & 10.99 & 10.88 & 16.69 & 16.80 & 21.17 & 21.20 \\
        \midrule
        & \clean & 38.48 & 64.33 & 5.90 & 16.10 & 35.73 & 72.26 & 9.90 & 21.20 & 4.90 & 17.40 & 24.20 & 48.40 & 4.97 & 12.82 & 10.62 & 28.63 & 16.84 & 35.14 \\ 
        & \asr & 95.39 & 97.49 & 98.80 & 99.30 & 97.26 & 98.86 & 94.70 & 96.90 & 98.80 & 99.90 & 99.20 & 100.0 & 97.22 & 98.81 & 97.90 & 99.40 & 97.41 & 98.83 \\
        & \modi & 236.42 & 234.41 & 214.09 & 214.44 & 201.71 & 202.06 & 226.78 & 226.96 & 238.13 & 236.93 & 133.35 & 133.06 & 195.75 & 194.03 & 524.57 & 522.88 & 246.35 & 245.60 \\
        \multirow{-4}{*}{\textsf{\textbf{@2}}} & \modn & 27.12 & 27.0 & 16.87 & 16.92 & 25.61 & 25.81 & 41.67 & 41.79 & 13.69 & 13.66 & 16.75 & 16.73 & 11.01 & 10.91 & 16.67 & 16.79 & 21.17 & 21.20 \\
        \midrule
        & \clean & 13.23 & 27.15 & 0.80 & 3.40 & 9.02 & 34.02 & 3.0 & 6.80 & 1.50 & 3.50 & 1.40 & 12.0 & 0.89 & 2.49 & 1.20 & 3.90 & 3.88 & 16.66 \\ 
        & \asr & 91.58 & 95.39 & 97.0 & 98.80 & 94.52 & 97.60 & 93.10 & 95.90 & 97.40 & 99.30 & 94.80 & 99.10 & 95.83 & 98.11 & 95.89 & 99.20 & 95.07 & 97.92 \\
        & \modi & 239.26 & 237.11 & 212.59 & 213.98 & 202.82 & 202.18 & 226.88 & 226.92 & 239.68 & 237.66 & 135.64 & 133.94 & 197.18 & 195.07 & 523.93 & 523.35 & 247.18 & 246.28 \\
        \multirow{-4}{*}{\textsf{\textbf{@3}}} & \modn & 27.28 & 27.11 & 16.93 & 16.88 & 25.26 & 25.62 & 41.59 & 41.72 & 13.69 & 13.68 & 16.96 & 16.84 & 11.05 & 10.97 & 16.71 & 16.75 & 21.18 & 21.20 \\
        \midrule
        & \clean & 1.20 & 3.01 & 0 & 0 & 0 & 3.08 & 0 & 0 & 0 & 0 & 0 & 0 & 0 & 0 & 0 & 0 & 0.15 & 0.76 \\ 
        & \asr & 88.28 & 94.09 & 94.10 & 98.20 & 91.10 & 96.92 & 91.60 & 94.70 & 97.40 & 99.10 & 90.60 & 97.50 & 94.04 & 97.51 & 94.99 & 99.0 & 92.76 & 97.13 \\
        & \modi & 241.56 & 239.12 & 212.66 & 214.32 & 114.09 & 202.45 & 226.61 & 226.92 & 239.68 & 237.94 & 137.80 & 135.35 & 199.09 & 195.80 & 525.71 & 523.85 & 248.23 & 246.97 \\
        \multirow{-4}{*}{\textsf{\textbf{@4}}} & \modn & 27.73 & 27.28 & 17.18 & 16.92 & 5.52 & 25.52 & 41.52 & 41.66 & 13.70 & 13.68 & 17.22 & 17.0 & 11.17 & 11.01 & 16.76 & 16.75 & 21.30 & 21.23 \\
        \bottomrule
    \end{tabular}}

\vspace{-0.3cm}
\label{tab:Untargeted_K10}
\end{table*}

\subsubsection*{Successful Attacks}\label{sec:success}
According to the definitions provided in Section~\ref{sec:problem-definition}, in the \textbf{targeted} scenario, we deem an attack as successful when at least $i$ variants in \variants are among the \topk results when \fadv is used as a query over \pool; contrarily, in the \textbf{untargeted} case, an adversarial example is successful when at least $i$ variants in \variants are ranked outside the \topk when \fadv is used as query over \pool.

As seen in Section~\ref{sec:dataset}, each attack considers a set \variants of exactly 4 variants, meaning $i \le 4$. The attack becomes more challenging as $i$ increases. For instance, when $i=1$, a targeted attack is successful when just one variant is ranked among the \topk. Conversely, when $i=4$, all variants must be in the \topk, making the attack harder.

Also, the factor $K$ affects the outcome of the attack, specifically low values of $K$ make the targeted attack harder. For example, let's assume $i=4$; with $K=4$ the attack succeeds only if the \topk set is exactly the set of variants. When increasing $K$ to 5, the attack is successful even if one function of \pool not in \variants is in the \topk. Conversely, increasing $K$ makes the untargeted attack more challenging. For instance, when $K=|P|-4$ the attack succeeds only when the \topk equals $\mPool \setminus \mV$.

The size of \pool impacts the result of the attack. As the pool size increases, a targeted attack becomes more difficult because more functions could be ranked among the \topk. Contrarily, in the untargeted case, the attack becomes more difficult as the pool size decreases, since fewer functions in \pool can be ranked within the \topk.

\subsubsection*{Performance Metrics}\label{sec:metrics}
We evaluate the robustness of the target models by using the Attack Success Rate (\asr) as the main metric, which is the percentage of adversarial examples that meet the success condition. We calculate the \asr for each value of $i \in \{1, 4\}$ (\asrone, \asrtwo, \asrthree, \asrfour). Recall that in \asri we consider successful the targeted attacks that bring at least $i$ variants in \topk. In the untargeted case, the attacks have to move at least $i$ variants outside \topk.

We also defined an aggregated \asr, namely \wasr, in which we decreasingly penalize the success of an experiment depending on the number of variants satisfying the success condition. For example, if when using \fadv as query only one of the variants is among the \topk, then this experiment will count for 0.25, if there are two it will count for 0.50, and so on.

We use the percentage of query functions where the success condition at $i$ is met before executing the attack, namely \cleani metrics for $i \in \{1, 4\}$ as reference. 

% Similarly to the \asr, to assess the performance of the target models on clean data, we define different \cleani metrics for $i \in \{1, 4\}$. For the targeted case, \cleani represents the percentage of cases for which at least $i$ variants in \variants are \textbf{not} ranked in the \topk. For the untargeted case, \cleani represents the percentage of cases for which at most $i$ variants in \variants are ranked in the \topk.
In our untargeted attack experiments, we also compute the standard \recallK before (\crecall) and after (\arecall) the attack. This quantifies the model's performance on clean data and its degradation following untargeted attacks.

We further investigate the quality of our attack using two other support metrics for each value of $i \in \{1, 4\}$, computed over the set of successful examples. The first metric, \textbf{M-Instrs@\textit{i}}, represents the number of new instructions in \fadv at the end of the attack. The second metric, \textbf{M-Nodes@\textit{i}}, measures the number of new nodes in \fadv at the end of the attack.

%\begin{itemize}
%    \item \textbf{Number of inserted instructions} (\textbf{M-Instrs@\textit{i}}): number of new instructions in $f_{adv}$ at the end of the attack;
%    \item \textbf{Number of inserted nodes} (\textbf{M-Nodes@\textit{i}}): number of new nodes in $f_{adv}$ at the end of the attack;
%\end{itemize}

\subsubsection*{Parameters of the Attack}\label{sec:parameters}
We run our attack for up to 30 iterations, exploring $\lambda$ values of 0, 0.01, and 0.3 (see Section~\ref{sec:objective}) to assess its impact on our results. Given that the average length of our query functions is approximately 100, we perturb 50 positions—about half of the total.
As in~\cite{jTrans-ISSTA22}, we use pool sizes of $32, 128, 512,$ and $1000$. 

Due to the computational overhead of dynamically updating the set of candidates (see Section~~\ref{sec:semPres}), the \textbf{SA} and \textbf{DBA} transformations are significantly slower than the others. To keep our experiment durations reasonable, we limit these transformations by testing 20 strands from a candidates' set composed by 100 strands with 50\% of random strands.

%We run our attack up to 30 iterations. 
%To assess the impact of different values of $\lambda$ (see Section~\ref{sec:objective}) on the experimental results, we consider three values for this parameter: \{0, 0.01, 0.3\}.

%To determine how many positions to perturb, we first calculate the average length of our query functions, which is about 100. Based on this, we select 50 positions, representing roughly half of the total. 

%The \textbf{SA} and the \textbf{DBA} transformations take significantly more time to execute compared to the others, primarily due to the dynamic updating of the candidates' set discussed in Section~\ref{sec:semPres}. To keep the duration of our experiments reasonable, we limit the frequency of these transformations during each iteration of the greedy optimizer. We then applied these transformations only to a fixed percentage of positions, specifically 40\% of the total. We also set the size of the candidates' set to 100 and fixed to 0.5 the percentage of random strands within the set.

%Following the evaluation performed in~\cite{jTrans-ISSTA22}, we consider pools having size \{32, 128, 512, 1000\}.

\subsection{RQ1: Targeted vs Untargeted Attacks}\label{sec:rq1}
In this section, we investigate the robustness of the considered models when subject to \textbf{targeted} and \textbf{untargeted} attacks. For brevity, we report in the tables and the plots only the results corresponding to the 128 and 512 pools and to $\lambda = 0$, which is the worst case according to the modification size. Furthermore, we discuss only the results @4, corresponding to the more difficult setup, and the \wasr. For the complete results, see Appendix~\ref{sec:app_A}.

\begin{figure}[h!]
    \centering
    \includegraphics[width=3.3in, trim = 0cm 0cm 0cm 0cm]{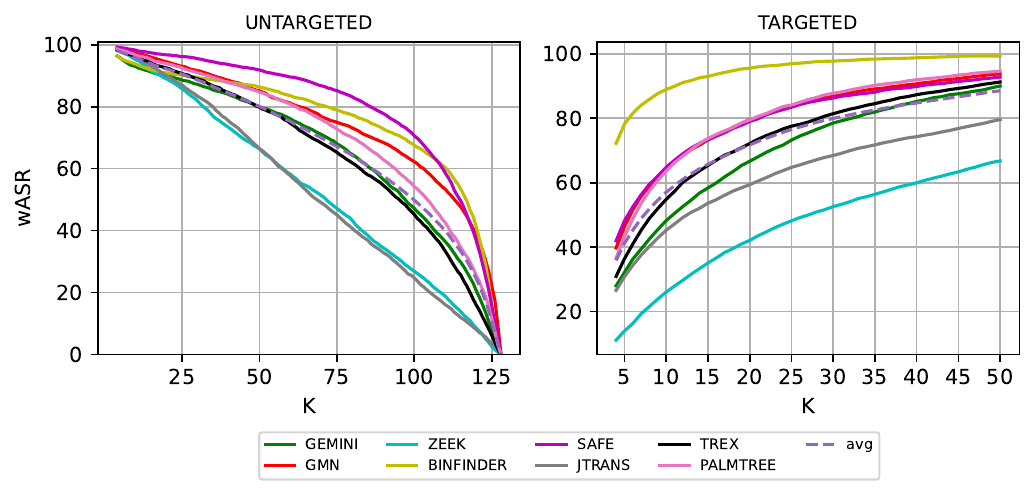}
    \caption{\wasr while varying the $K$ value and considering $|\mPool| = 128$. The dotted curve represents the average values across the different models.}
    \label{img:asr_k_tar_untar_128}
    \vspace{-0.3cm}
\end{figure}

\begin{table*}
\centering
\caption{Targeted attack at $K=10$ when considering pools with size 128 and 512 and setting $\lambda = 0$. In column AVG we report the average of the measures across all models.}
\resizebox{\textwidth}{!}{%
    \begin{tabular}{|c|c|rr|rr|rr|rr|rr|rr|rr|rr|rr|} 
        \cmidrule[\heavyrulewidth]{3-18}
        \multicolumn{1}{l}{\multirow{3}{*}{}} & \multicolumn{1}{l|}{\multirow{2}{*}{}} & \multicolumn{16}{c|}{\textbf{Models}} & \multicolumn{2}{c}{\multirow{1}{*}{}} \\
        \cmidrule{3-20}
        \multicolumn{1}{l}{} & \multicolumn{1}{l|}{} & \multicolumn{2}{c|}{\gemini} & \multicolumn{2}{c|}{\gmn} & \multicolumn{2}{c|}{\zeek} & \multicolumn{2}{c|}{\binfinder} & \multicolumn{2}{c|}{\safe} & \multicolumn{2}{c|}{\jtrans} & \multicolumn{2}{c|}{\trex} & \multicolumn{2}{c|}{\palmtree} & \multicolumn{2}{c|}{AVG} \\
        \cmidrule{2-20}
        \multicolumn{1}{l|}{} & \bm{$|\mPool|$} & \textbf{128} & \textbf{512} & \textbf{128} & \textbf{512} & \textbf{128} & \textbf{512} & \textbf{128} & \textbf{512} & \textbf{128} & \textbf{512} & \textbf{128} & \textbf{512} & \textbf{128} & \textbf{512} & \textbf{128} & \textbf{512} & \textbf{128} & \textbf{512} \\
        \toprule
        \toprule
        & \wasr & 48.25 & 25.10 & 64.68 & 40.15 & 26.07 & 9.34 & 88.95 & 71.37 & 64.80 & 40.02 & 45.25 & 23.35 & 54.85 & 30.48 & 63.60 & 30.24 & 57.06 & 33.76 \\
        \toprule
        \toprule
        & \clean & 17.20 & 5.70 & 15.80 & 3.90 & 15.0 & 3.47 & 12.12 & 3.11 & 13.90 & 4.0 & 23.90 & 4.80 & 12.10 & 3.30 & 14.97 & 4.11 & 15.62 & 4.05 \\ 
        & \asr & 71.90 & 45.10 & 78.60 & 59.80 & 43.16 & 19.49 & 94.39 & 80.96 & 80.30 & 59.10 & 80.20 & 49.10 & 68.80 & 44.60 & 78.04 & 47.19 & 74.42 & 50.67 \\
        & \modi & 173.62 & 170.18 & 214.53 & 210.35 & 94.86 & 101.21 & 155.21 & 154.78 & 123.74 & 128.05 & 77.29 & 84.46 & 112.70 & 112.76 & 193.41 & 182.05 & 143.17 & 142.98 \\
        \multirow{-4}{*}{\textsf{\textbf{@1}}} & \modn & 13.83 & 15.23 & 18.87 & 18.35 & 11.18 & 10.80 & 33.28 & 33.0 & 15.72 & 15.72 & 11.95 & 12.66 & 11.44 & 11.36 & 18.28 & 18.62 & 16.82 & 16.97 \\
        \midrule
        & \clean & 9.40 & 2.30 & 8.10 & 1.90 & 6.53 & 0.71 & 7.82 & 1.80 & 8.60 & 1.40 & 8.30 & 0.80 & 8.20 & 1.60 & 7.98 & 1.40 & 8.12 & 1.49 \\ 
        & \asr & 57.0 & 29.20 & 71.10 & 47.30 & 31.94 & 11.33 & 91.68 & 75.55 & 71.40 & 46.30 & 58.90 & 28.0 & 60.40 & 34.40 & 70.06 & 34.97 & 64.06 & 38.38 \\
        & \modi & 173.68 & 173.84 & 213.81 & 214.23 & 96.81 & 108.93 & 155.94 & 156.73 & 126.32 & 128.05 & 84.68 & 94.18 & 112.83 & 114.63 & 192.08 & 173.26 & 144.52 & 145.67 \\
        \multirow{-4}{*}{\textsf{\textbf{@2}}} & \modn & 14.68 & 16.26 & 18.63 & 18.68 & 10.35 & 11.50 & 33.40 & 33.46 & 15.89 & 15.72 & 12.65 & 13.61 & 11.30 & 11.13 & 18.09 & 18.08 & 16.87 & 17.30 \\
        \midrule
        & \clean & 2.70 & 0.50 & 3.30 & 0.40 & 1.84 & 0 & 4.91 & 0.80 & 3.10 & 0.3 & 1.30 & 0.1 & 4.10 & 0.50 & 2.89 & 0.10 & 3.02 & 0.34 \\ 
        & \asr & 38.0 & 15.70 & 60.40 & 32.30 & 18.57 & 4.29 & 87.37 & 67.33 & 60.10 & 32.30 & 27.30 & 10.40 & 49.50 & 24.10 & 57.58 & 22.14 & 49.85 & 26.07 \\
        & \modi & 177.87 & 181.08 & 214.63 & 224.91 & 93.72 & 108.83 & 157.19 & 159.0 & 129.75 & 131.05 & 96.62 & 105.67 & 115.36 & 114.20 & 189.60 & 175.48 & 146.84 & 150.01 \\
        \multirow{-4}{*}{\textsf{\textbf{@3}}} & \modn & 15.76 & 17.45 & 18.92 & 19.17 & 10.26 & 11.14 & 33.67 & 33.88 & 15.84 & 15.88 & 13.86 & 15.56 & 11.34 & 11.20 & 18.15 & 19.18 & 17.23 & 17.93 \\
        \midrule
        & \clean &  0.90 & 0.20 & 1.60 & 0 & 0.31 & 0 & 3.21 & 0.40 & 1.40 & 0 & 0.30 & 0.10 & 2.60 & 0.30 & 1.20 & 0 & 1.44 & 0.13 \\
        & \asr & 26.10 & 10.40 & 48.60 & 21.20 & 10.61 & 2.24 & 82.36 & 61.62 & 47.40 & 22.40 & 14.60 & 5.9 & 40.70 & 18.80 & 48.70 & 16.63 & 39.88 & 19.90 \\
        & \modi & 190.36 & 192.19 & 219.14 & 239.85 & 104.95 & 124.5 & 159.24 & 160.23 & 136.29 & 140.71 & 109.49 & 118.15 & 116.21 & 117.37 & 192.21 & 176.20 & 153.49 & 158.65 \\
        \multirow{-4}{*}{\textsf{\textbf{@4}}} & \modn & 16.57 & 18.25 & 19.48 & 19.49 & 10.99 & 13.45 & 34.19 & 34.12 & 16.05 & 16.40 & 14.71 & 16.24 & 11.42 & 11.53 & 18.37 & 19.39 & 17.72 & 18.60 \\
        \bottomrule
    \end{tabular}}
\vspace{-0.3cm}
\label{tab:Targeted_K10}
\end{table*}

\subsubsection{Untargeted Attacks}\label{sec:rq1_untargeted}

Table~\ref{tab:Untargeted_K10} presents the results for the \textbf{untargeted} attack scenario.

For $|\mPool| = 128$, we observe an average \cleanfour of 0.15\%, indicating that the success condition is virtually never met without the attack. After applying the attack and querying the pool with the generated \fadv, we achieve an average \asrfour of 92.76\%, meaning that in over 9 out of 10 cases, all variants in \variants are pushed outside the Top-10. Furthermore, when considering the \wasr, the attack succeeds in 95.81\% of cases.

Figure~\ref{img:asr_k_tar_untar_128} shows the average \wasr for $K$ values ranging from 4 to 128. For small $K$ values ($\leq 25$), the average \wasr remains well above 90\% but decreases sharply as the search depth increases, consistent with observations in Section~\ref{sec:success}. However, see Figure~\ref{img:asr_k_tar_untar_512}, the \wasr maintains an average consistently above 80\%, indicating that a larger difference between $K$ and the pool size significantly benefits the attacker. Notably, when the binary function similarity model is used to detect vulnerable or malicious functions, the pool likely contains thousands of functions, making untargeted attacks significantly easier.

The \recall metric further underscores the models' vulnerability to untargeted attacks. Specifically, for a pool of 128 functions, the \recall decreases from an initial average of 0.86 to 0.04 after running the attack.

%all models show an initial top performance on clean data, with an average \recall of 0.86, which drops to 0.04 after running the attack.

%Our initial claim is confirmed; almost all the models show initial top performance on clean data, as demonstrated by the average initial \recall of 0.86, while they are not robust against our adversarial examples, as confirmed by the average \wasr of 95.81\%.

%while our attack is successful in more than 9 out of 10 cases, with a \wasr of 95.81\%.

\subsubsection{Targeted Attacks}\label{sec:rq1_targeted}
Table~\ref{tab:Targeted_K10} presents the attack results for the \textbf{targeted} scenario. With a pool of 128 functions, the average \cleanfour is 1.44\%, indicating that variants in \variants are ranked in the Top-10 for their corresponding \fq in less than 2\% of cases. The average \asrfour is 39.88\%, meaning that when querying with \fadv, all variants in \variants are ranked in the Top-10 nearly 40\% of the time. The \wasr is 57.06\%.

\begin{figure}[h!]
    \centering
    \includegraphics[width=3.3in, trim = 0cm 0cm 0cm 0cm]{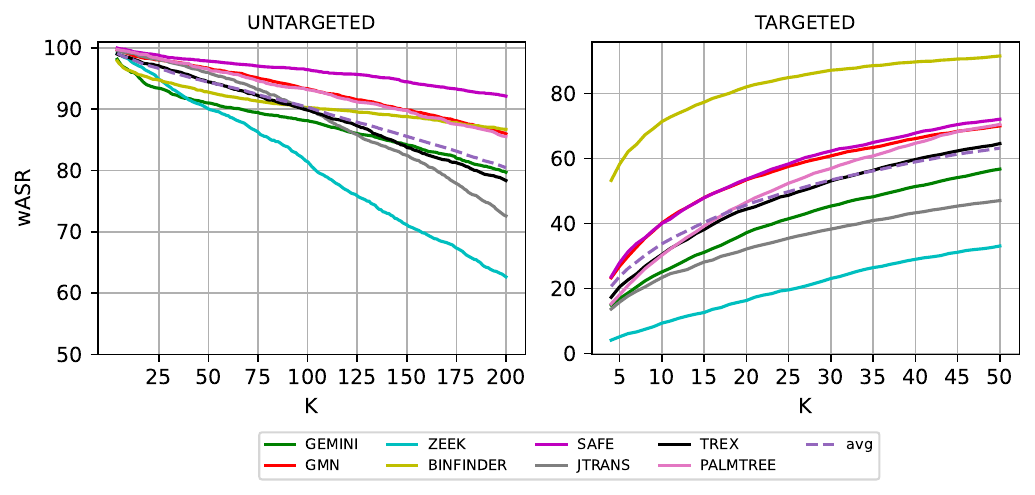}
    \caption{\wasr while varying the $K$ value and considering $|\mPool| = 512$. The dotted curve represents the average values across the different models.}
    \label{img:asr_k_tar_untar_512}
    \vspace{-0.3cm}
\end{figure}

Figure~\ref{img:asr_k_tar_untar_128} illustrates the average \wasr for targeted attacks across $K \in [4, 50]$ when querying a pool of 128 functions, showing a rapid increase to values exceeding 40\%. Figure~\ref{img:asr_k_tar_untar_512} depicts similar trends with a pool of 512 functions; however, the \wasr increases more gradually compared to $|\mPool| = 128$. As reported in Appendix~\ref{sec:app_A}, model robustness further increases with a pool of 1000 functions, yielding an average \wasr of 25.25\%.

\begin{table*}[]
\centering
\caption{Transferability matrix for the untargeted attack case, considering $|\mPool|=128$ and $K=10$. In the rows, we indicate the model for which the adversarial examples were created, and in the columns, the model on which the examples are tested. Each value represents the \wasr.}
\resizebox{0.8\textwidth}{!}{%
\begin{tabular}{|c|rrrrrrrr|r|}
    \cmidrule{2-10}
    %\multicolumn{1}{c|}{} & \multicolumn{1}{c}{\gemini} & \multicolumn{1}{c}{\gmn} & \multicolumn{1}{c}{\zeek} & \multicolumn{1}{c}{\binfinder} & \multicolumn{1}{c}{\safe} & \multicolumn{1}{c}{\jtrans} & \multicolumn{1}{c}{\trex} & \multicolumn{1}{c|}{\palmtree} & \multicolumn{1}{c|}{Row AVG} \\
    \multicolumn{1}{c|}{} & \gemini & \gmn & \zeek & \binfinder & \safe & \jtrans & \trex & \palmtree & \trate \\
    \cmidrule{1-10}
    \gemini & $\blacksquare$ & 70.67 & 82.34 & 23.55 & 44.69 & 42.66 & 35.70 & 71.92 & 53.08 \\
    \gmn & 68.30 & $\blacksquare$ & 85.02 & 20.30 & 50.78 & 50.98 & 47.77 & 67.17 & 55.76 \\
    \zeek & 71.59 & 76.13 & $\blacksquare$ & 20.08 & 62.40 & 56.58 & 60.39 & 70.91 & 59.73 \\
    \binfinder & 79.90 & 80.15 & 85.65 & $\blacksquare$ & 60.68 & 61.18 & 58.77 & 68.19 & 70.65 \\
    \safe & 61.02 & 62.38 & 84.38 & 24.22 & $\blacksquare$ & 72.12 & 79.57 & 68.77 & 64.64 \\
    \jtrans & 63.60 & 57.38 & 77.38 & 20.15 & 65.05 & $\blacksquare$ & 60.60 & 59.43 & 57.66 \\
    \trex & 58.02 & 57.99 & 82.36 & 19.96 & 76.95 & 67.69 & $\blacksquare$ & 71.44 & 62.06 \\
    \palmtree & 71.39 & 77.15 & 87.27 & 31.41 & 83.74 & 74.15 & 88.33 & $\blacksquare$ & 73.35 \\
    \cmidrule{1-10}
    \vrate & 67.69 & 68.84 & 83.49 & 22.81 & 63.47 & 60.77 & 61.59 & 68.26 & $\blacksquare$ \\
    \hline
    \hline
    \textit{random} & $67.81 \pm 0.06$ & $55.68 \pm 0.73$ & $79.92 \pm 0.15$ & $17.87 \pm 0.97$ & $45.94 \pm 0.30$ & $49.69 \pm 0.80$ & $31.66 \pm 1.02$ & $53.16 \pm 0.32$ & 50.22 \\
    \bottomrule
\end{tabular}}
\vspace{-0.3cm}
\label{tab:Untar_Transf_K10}
\end{table*}

\subsubsection{Impacts of the Modification Size}\label{sec:lambda}
We study how the modification size impacts models robustness by varying the $\lambda$ parameter in the Equation~\ref{eq:obj}. We analyzed three models selected to be representative of all DNN architectures (specifically, \gemini, \safe, and \jtrans) with $\lambda$ values of $0$, $0.01$, and $0.3$. The complete results are reported in Appendix~\ref{sec:app_A}.

As shown in Figure~\ref{img:asr_k_tar_untar_128_0_03}, increasing $\lambda$ leads to a significant decrease in the average \wasr\ in both untargeted and targeted scenarios. In the untargeted case, the \wasr goes from 95.76\% at $\lambda=0$ to 77.39\% at $\lambda=0.01$, and to 24.59\% at $\lambda=0.3$.  Correspondingly, the average number of modifications, \modi\textbf{@4}, decreases from 206.34 at $\lambda=0$ to 32.26 at $\lambda=0.01$, and to 20.26 at $\lambda=0.3$. In the targeted case, the average \wasr goes from 52.77\% at $\lambda=0$ to 27.74\% at $\lambda=0.01$, and to 9.94\% at $\lambda=0.3$, while the \modi\textbf{@4} decreases from 145.38 at $\lambda=0$ to 19.10 at $\lambda=0.01$, and to 6.19 at $\lambda=0.3$.

\begin{figure}[h!]
    \centering
    \includegraphics[width=3.3in, trim = 0cm 0cm 0cm 0cm]{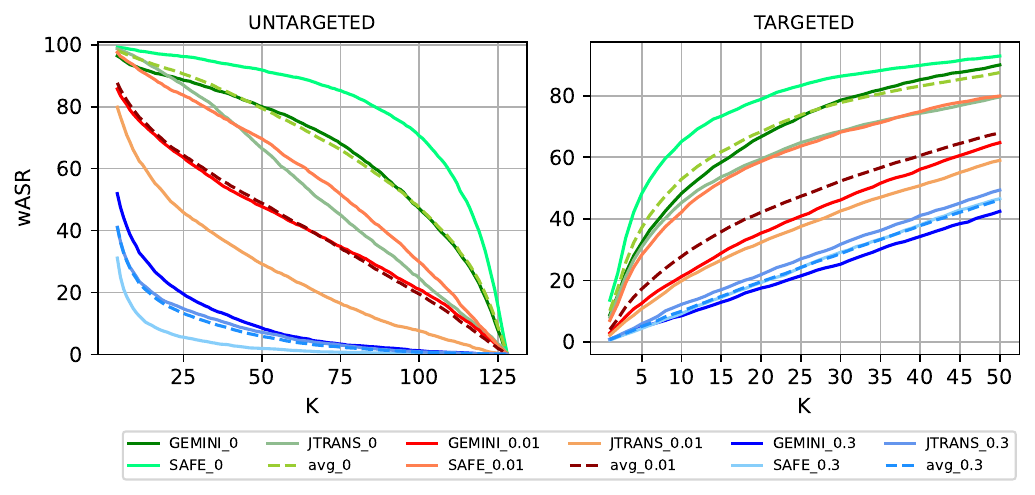}
    \caption{\wasr while varying the $K$ value, considering $|\mPool| = 128$  and $\lambda \in \{0, 0.01, 0.3\}$. For each value of $\lambda$, we show the \wasr together with the average on three models differing in architecture. We represent in \color{mygreen}\textbf{green} \color{black} the results corresponding to $\lambda=0$, in \color{red}\textbf{red} \color{black} the results corresponding to $\lambda=0.1$, and in \color{blue}\textbf{blue} \color{black} the ones for $\lambda=0.3$.}
    \label{img:asr_k_tar_untar_128_0_03}
    \vspace{-0.3cm}
\end{figure}

%the average \asrfour over the considered models is 2.16, with an average of 5.48 instructions and 0.89 nodes inserted (details in Appendix~\ref{sec:app_A}). T

\begin{mybox}
\textbf{Key takeaway.} All models are more susceptible to untargeted attacks than targeted ones. Specifically, the average \wasr is 95.81\% vs 57.06\% when querying a pool of 128 functions, and 98.28\% vs 33.76\% when querying a pool of 512 functions.
\end{mybox}

\subsection{RQ2: Generalizability and Transferability}\label{sec:rq2}

In this section, we discuss the generalizability of our approach, showing how the considered metrics vary across different models. We investigate whether the generated adversarial examples can be transferred across diverse models, to demonstrate the variation in the attack success rate when adversarial examples intended for one model are presented to a different model.
% [67.77, 67.77, 67.90], [56.23, 54.65, 56.15], [80.0, 80.05, 79.70], [18.57, 16.50, 18.55], [46.15, 45.52, 46.15], [49.12, 50.82, 49.12], [30.98, 33.10, 30.90], [52.83, 53.60, 53.05]

\subsubsection{Generalizability}\label{sec:rq2_gen}

Our black-box transformations alter both the CFG and instruction sequence of a function. A key point is whether they are sufficient to attack all examined models.

In the untargeted setting with $|\mPool|=128$, Figure~\ref{img:asr_k_tar_untar_128} shows similar performance across most models at various levels of $K$, especially for $K<25$, as confirmed in Table~\ref{tab:Untargeted_K10}. Here, \gemini is the most robust model (\wasr = 92.99\%), while \safe is the weakest (\wasr = 98.30\%).

%Figure~\ref{img:asr_k_tar_untar_128} shows the results for the untargeted scenario with $|\mPool|=128$, indicating that the performance at various levels of $K$ is comparable across most models. Specifically, for values of $K$ less than 25, all models behave virtually the same, as reported in Table~\ref{tab:Untargeted_K10}. In this table, \gemini exhibits greater robustness with a \wasr of 92.99\%, while \safe is the least robust model with a \wasr of 98.30\%.

As visible in Figure~\ref{img:asr_k_tar_untar_512}, this trend holds for larger pools. As $K$ increases, \zeek and \jtrans stand out as the most robust models, with \safe remaining the weakest. Even at $K=100$, \zeek is fooled in over 80\% of cases, suggesting that blacklist defenses implemented by binary function similarity models can be bypassed in practice (e.g., a vulnerable function ranked beyond the top 100 may be ignored). We speculate that \safe is more vulnerable due to its reliance on pre-trained embeddings of preprocessed instructions, making it sensitive to transformations like \textbf{SA} and \textbf{DBA}. 

Interestingly, Table~\ref{tab:Untargeted_K10} shows that both \zeek and \gemini have some of the lowest \crecall values among the models, yet their \arecall values are among the highest, meaning they are more robust against our attack. In contrast, models with top performance on clean data, like \gmn, \safe, and \trex, are the ones exhibiting less robustness, as they show lower \arecall values.

As visible in Table~\ref{tab:Targeted_K10} and in Figure~\ref{img:asr_k_tar_untar_128}, the performance at $K=10$ in the \textbf{targeted} scenario is comparable across the majority of the models, except for \binfinder, which is the least robust model, and \zeek which is the most robust one.  \binfinder learns function semantics considering features like VEX-IR instructions and constants, which are heavily impacted by \textbf{SA} and \textbf{DBA}. All the other models exhibit consistent robustness, with \jtrans being the most robust and \safe the least, showing a \wasr of 45.25\% and 64.80\% respectively. Overall, the models relying on an instructions-based representation (such as \binfinder and \safe) seem to be less robust against our attack. As visible in Figure~\ref{img:asr_k_tar_untar_512}, this same analysis holds when increasing the pool size to 512.

\begin{mybox}
\textbf{Key takeaway.} In the untargeted case, the performance at various levels of $K$ is comparable across most models, with \safe being the least robust and \zeek the most robust model. While in the targeted case, \zeek is the most robust model and \binfinder the least robust.

Top performance on clean data does not correspond to greater robustness. Models showing greater \crecall, often exhibit lower \arecall, whereas those with poorer clean data performance, tend to have higher \arecall.
\end{mybox}

\begin{table*}[]
\centering
\caption{Transferability matrix for the targeted attack case, considering $|\mPool|=128$ and $K=10$. In the rows, we indicate the model for which the adversarial examples were created, and in the columns, the model on which the examples are tested. Each value represents the \wasr.}
\resizebox{0.8\textwidth}{!}{%
\begin{tabular}{|c|rrrrrrrr|r|}
    \cmidrule{2-10}
    \multicolumn{1}{c|}{} & \gemini & \gmn & \zeek & \binfinder & \safe & \jtrans & \trex & \palmtree & \trate \\
    \cmidrule{1-10}
    \gemini & $\blacksquare$ & 10.80 & 10.05 & 7.62 & 9.03 & 10.10 & 8.08 & 8.04 & 9.10 \\
    \gmn & 10.80 & $\blacksquare$ & 9.57 & 7.83 & 11.72 & 11.05 & 7.92 & 10.30 & 9.88 \\
    \zeek & 6.35 & 7.78 & $\blacksquare$ & 6.93 & 10.49 & 11.50 & 6.68 & 6.56 & 8.04 \\
    \binfinder & 5.61 & 5.96 & 9.52 & $\blacksquare$ & 9.57 & 10.02 & 7.79 & 6.38 & 7.84 \\
    \safe & 8.65 & 9.93 & 10.47 & 8.10 & $\blacksquare$ & 12.0 & 12.28 & 9.92 & 10.19 \\
    \jtrans & 8.20 & 8.88 & 8.85 & 7.52 & 12.3 & $\blacksquare$ & 10.50 & 8.47 & 9.25 \\
    \trex & 7.70 & 9.53 & 8.97 & 7.47 & 16.70 & 13.60 & $\blacksquare$ & 11.14 & 10.73 \\
    \palmtree & 10.20 & 13.38 & 7.31 & 9.49 & 11.72 & 11.40 & 13.18 & $\blacksquare$ & 10.95 \\
    \cmidrule{1-10}
    \vrate & 8.22 & 9.47 & 9.25 & 7.85 & 11.65 & 11.38 & 9.49 & 8.69 & $\blacksquare$ \\
    \hline
    \hline
    \textit{random} & $6.65 \pm 0.11$ & $6.73 \pm 0.22$ & $8.67 \pm 0.24$ & $7.26 \pm 0.03$ & $8.57 \pm 0.09$ & $9.89 \pm 0.10$ & $6.38 \pm 0.32$ & $6.09 \pm 0.08$ & 7.53 \\
    \bottomrule
\end{tabular}}
\vspace{-0.3cm}
\label{tab:Tar_Transf_K10}
\end{table*}

\subsubsection{Transferability}\label{sec:rq2_transf}
Our attack strategy employs a greedy optimizer that iteratively applies transformations based on feedback from the target model. An intriguing aspect to explore is whether an adversarial example generated against one model can be leveraged to target all the other models under analysis. We define this property as the Transferability Success Rate (\trate). Furthermore, we want to investigate how models react against adversarial examples designed for other models. We define this property as the Vulnerability Rate (\vrate).

To evaluate these two properties, we compare using a simple baseline that applies a sequence of random transformations to the query function \fq. The random baseline is run for the same number of iterations as the greedy procedure, and the experiment is repeated three times. Note that this baseline can be used to compare a random application of transformations against our optimizer, see the difference between these results and the one in Section~\ref{sec:rq1}.

Table~\ref{tab:Untar_Transf_K10} presents the results in terms of \wasr of the transferability experiment in the untargeted case, together with the average \wasr and the standard deviation for the random baseline. Table~\ref{tab:Tar_Transf_K10} presents the results for the transferability experiment in the targeted scenario.

\subsubsection*{Transferability Success Rate}

When considering the untargeted scenario, the \trate values indicate that adversarial examples generated against \palmtree and \binfinder are the most transferable to other models, with \trate values of 73.35\% and 70.65\% respectively. We attribute these results to the transformations applied when attacking \palmtree and \binfinder. In these cases, \textbf{SA} and \textbf{DBA} are the most used transformations. These two modify most of the features considered by the target models (namely, the content and the topology of the CFG). We will further discuss this aspect in the next section where we will peruse the frequency of transformations applied against each model. Finally, all the transferred examples demonstrate higher effectiveness when compared to the random baseline, with an average \trate of 62.12\% vs 50.22\% respectively.

In the targeted scenario, the \trate results show that adversarial examples transfer less effectively compared to those from the untargeted case. This is expected, given that targeted attacks are generally less effective than untargeted ones, as discussed in Section~\ref{sec:rq1}. Nevertheless, the transferred examples outperform those produced by the random baseline. Finally, similar to the untargeted case, adversarial examples generated against \palmtree remain the most effective, with a \trate of 10.95\%.

\subsubsection*{Vulnerability Rate}
Interestingly, the \vrate results show that \binfinder stands out as the most robust model, with a lower \vrate of 22.81\%. We believe that adversarial examples targeting \binfinder must possess unique features that are not present in those generated against other models. We will further discuss this insight in the next section. Surprisingly, \zeek is the least robust model against transferred adversarial examples, with a \vrate value of 83.49\%. Generally, transferred examples are more effective than those generated using the random baseline, except \gemini, where the corresponding \vrate value is comparable to the \wasr achieved with the random baseline.

In the targeted scenario, all models perform similarly, with \binfinder being the most robust (\vrate of 7.85\%) and \safe the least (\vrate of 11.65\%). Despite these low values, transferred examples still outperform the random baseline.

\begin{mybox}
\textbf{Key takeaway.} The transferability of adversarial examples across models is more effective in the untargeted context than in the targeted scenario. Additionally, the distribution of applied transformations directly affects the success of transferring an adversarial example to another model.
\end{mybox}

\subsection{RQ3: Common Model Behaviors}\label{sec:rq3}

In this section, we discuss whether or not an adversarial example can reveal common behaviors that the model applies when analyzing binary functions.

\begin{table*}[]
\centering
\caption{Untargeted and Targeted attack at $K=10$ with transformations applied individually, considering $|\mPool|=128$ and $\lambda = 0$. The $\Delta\%$ value represents the percentage improvement in terms of \wasr achieved by considering all transformations (ALL) compared to applying each transformation in isolation.}
\resizebox{\textwidth}{!}{%
\begin{tabular}{|l|c|r|rrrr||r|rrrr||r|rrrr||r|rrrr|}
    \cmidrule[\heavyrulewidth]{3-17}
    \multicolumn{1}{l}{\multirow{3}{*}{}} & \multicolumn{1}{l|}{\multirow{3}{*}{}} & \multicolumn{15}{c|}{Models} & \multicolumn{5}{l}{} \\
    \cmidrule{3-22}
    \multicolumn{1}{l}{} & \multicolumn{1}{l|}{} & \multicolumn{5}{c||}{\gemini} & \multicolumn{5}{c||}{\safe} & \multicolumn{5}{c||}{\jtrans} & \multicolumn{5}{c|}{AVG} \\
    \cmidrule{3-22}
    \multicolumn{1}{l}{} & \multicolumn{1}{l|}{} & \multicolumn{1}{c|}{\textbf{ALL}} & \multicolumn{1}{c}{\textbf{IR}} & \multicolumn{1}{c}{\textbf{NS}} & \multicolumn{1}{c}{\textbf{DBA}} & \multicolumn{1}{c||}{\textbf{SA}} & \multicolumn{1}{c|}{\textbf{ALL}} & \multicolumn{1}{c}{\textbf{IR}} & \multicolumn{1}{c}{\textbf{NS}} & \multicolumn{1}{c}{\textbf{DBA}} & \multicolumn{1}{c||}{\textbf{SA}} & \multicolumn{1}{c|}{\textbf{ALL}} & \multicolumn{1}{c}{\textbf{IR}} & \multicolumn{1}{c}{\textbf{NS}} & \multicolumn{1}{c}{\textbf{DBA}} & \multicolumn{1}{c||}{\textbf{SA}} & \multicolumn{1}{c|}{\textbf{ALL}} & \multicolumn{1}{c}{\textbf{IR}} & \multicolumn{1}{c}{\textbf{NS}} & \multicolumn{1}{c}{\textbf{DBA}} & \multicolumn{1}{c|}{\textbf{SA}} \\
    \toprule
    \toprule
    \multicolumn{1}{|c||}{\multirow{3}{*}{\textbf{UNTARGETED}}} & \arecall & 0.07 & 0.71 & 0.32 & 0.07 & 0.07 & 0.02 & 0.93 & 0.79 & 0.27 & 0.01 & 0.04 & 0.85 & 0.57 & 0.22 & 0.02 & 0.04 & 0.83 & 0.56 & 0.19 & 0.03 \\
    \multicolumn{1}{|c||}{} & \wasr & 92.99 & 30.08 & 68.24 & 92.97 & 92.16 & 98.30 & 7.58 & 30.35 & 73.47 & 98.72 & 96.0 & 14.67 & 43.34 & 78.29 & 97.63 & 95.76 & 17.44 & 47.31 & 88.39 & 96.17 \\
    \multicolumn{1}{|c||}{} & $\Delta\%$ & $\blacksquare$ & 67.65 & 26.62 & 0.02 & 0.89 & $\blacksquare$ & 92.29 & 69.13 & 25.26 & -0.43 & $\blacksquare$ & 84.72 & 43.34 & 18.45 & -1.70 & $\blacksquare$ & 81.79 & 50.60 & 7.70 & -0.43 \\
    \midrule
    \multicolumn{1}{|c||}{\multirow{2}{*}{\textbf{TARGETED}}} & \wasr & 48.25 & 7.61 & 13.23 & 38.48 & 28.16 & 64.80 & 7.52 & 11.95 & 29.57 & 50.45 & 45.25 & 7.73 & 12.49 & 28.76 & 33.59 & 52.77 & 7.62 & 12.56 & 32.27 & 37.40 \\
    \multicolumn{1}{|c||}{} & $\Delta\%$ & $\blacksquare$ & 84.23 & 72.58 & 20.25 & 41.64 & $\blacksquare$ & 88.40 & 81.56 & 54.37 & 22.15 & $\blacksquare$ & 82.92 & 72.40 & 36.44 & 25.77 & $\blacksquare$ & 85.56 & 76.20 & 38.85 & 29.17 \\
    \bottomrule
\end{tabular}}
\vspace{-0.3cm}
\label{tab:isolated_K10}
\end{table*}

\subsubsection{Distribution of Applied Transformations}\label{sec:transf_distro}
\begin{figure}[h!]
    \centering
    \includegraphics[width=3.3in, trim = 0cm 0cm 0cm 0cm]{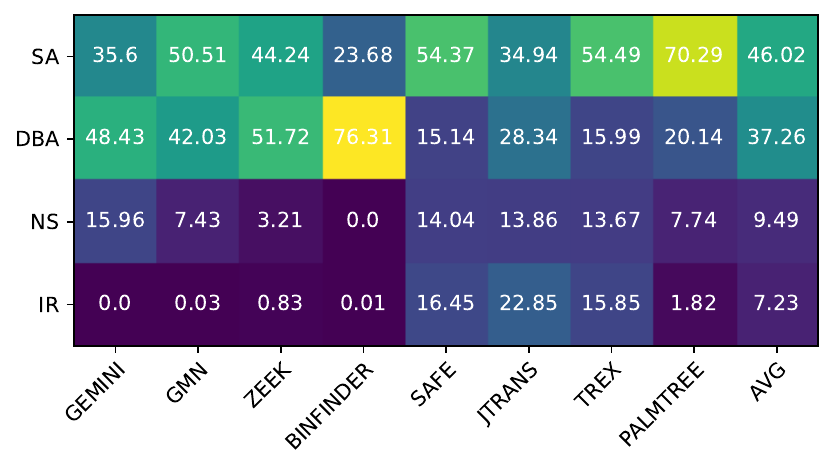}
    \caption{Distribution of applied transformations in the untargeted scenario, with $K=10, \lambda=0$ and querying a pool \pool of 128 functions. The AVG column shows the average distribution across the various models.}
    \label{img:Transf_perc_untar_128}
    \vspace{-0.1cm}
\end{figure}

Figure~\ref{img:Transf_perc_untar_128} shows the distribution of applied transformations across the various models together with the results in average in the untargeted scenario when querying a pool of 128 functions. The results are calculated by considering, for each model, only the adversarial examples that succeed according to the \asrfour metric.

Looking at the average results, it is evident that \textbf{SA} and \textbf{DBA} are the transformations that most contribute to the success of adversarial examples, being applied in 83.28\% of the time. In contrast, \textbf{IR}, an in-place transformation that neither alters the CFG nor introduces new instructions, is the least applied.

Our attack procedure can identify key aspects of the target model, particularly the target architecture and the function representation strategy.

As shown in Figure~\ref{img:Transf_perc_untar_128}, neither \gemini nor \gmn are affected by \textbf{IR}. This is reasonable, as both models utilize a GNN architecture that does not consider the position of the single instructions. In contrast, our strategy primarily focuses on transformations that either insert new nodes or new instructions. Specifically, when targeting \gemini, \textbf{DBA} is the most common choice, as it introduces both new instructions and new nodes into the CFG.

When moving to models that use instruction-based function representations (i.e., \zeek and \binfinder), \textbf{SA} and \textbf{DBA} become the most frequently applied transformations. This is due to both techniques adding new strands to the function's body.

\safe along with \jtrans and \trex, utilize architectures that consider the position of instructions within a function. As a result, \textbf{IR} is chosen more often compared to its usage in the previously mentioned models. It is interesting to note that the distribution of percentages against \jtrans is more balanced. This reflects the fact that \jtrans accounts for both the instructions and their positions within the function, as well as the CFG nodes, with \textit{jump} instructions being modeled differently from other instructions. 

Although \palmtree is built on an LSTM architecture similar to \safe, \textbf{SA} is chosen significantly more often in this case (70.29\% vs 54.37\%), whereas \textbf{IR} is rarely applied, unlike when attacking \safe. We believe this is due to the more sophisticated instruction embedding technique implemented by \palmtree, which causes our attack to focus more on transformations that insert new instructions rather than the ones manipulating the CFG or swapping existing instructions.

We emphasize that \textbf{SA} and \textbf{DBA} are universal transformations capable of modifying nearly all the features considered by the target models. This is evidenced not only by the previously discussed percentages but also by the intrinsic nature of these two transformations, which add new nodes and instructions to the modified function. Consequently, this explains why the adversarial examples generated against \palmtree and \binfinder transfer most effectively to the other models, as demonstrated by the \trate results discussed in the previous section.

\begin{mybox}
\textbf{Key takeaway.} The architecture of the target model and its function representation strategy significantly affect the type of transformation selected by our attack strategy. Specifically, transformations affecting the CFG are predominant when attacking models considering the CFG, while transformations inserting new instructions or altering their order are predominant when targeting models that do not consider the CFG topology.
\end{mybox}

\subsubsection{Transformations in Isolation}\label{sec:isol}
We now analyze the impact of individual transformations, focusing on three representative models—\gemini, \safe, and \jtrans—selected from the considered DNN architectures. For this evaluation, we run our greedy optimization strategy disabling all transformations but the one tested. To ensure a fair comparison, the number of candidates per iteration tested in this evaluation matches the number of candidates for the corresponding transformation in the main approach. For example, if transformation \textbf{IR} has 50 candidates per iteration in the main approach, the same number is used when evaluating \textbf{IR} alone.

Table~\ref{tab:isolated_K10} presents the results for the untargeted and targeted attacks performed considering the transformations in isolation. These confirm the findings from Section~\ref{sec:transf_distro}. Specifically, in the untargeted scenario, \textbf{DBA} and \textbf{SA}, the most frequent transformations for \gemini \safe, and \jtrans respectively, are also the most effective when used in isolation to target these models.  The $\Delta\%$ measure, which represents the percentage improvement in \wasr when combining transformations compared to applying each transformation individually, is $-0.43$ on \safe and $-1.70$ on \jtrans. This means that when considering \textbf{SA} alone, we can obtain results comparable to the main approach in terms of \wasr. The results in the targeted scenario confirm the effectiveness of \textbf{DBA} and \textbf{SA} in attacking the considered models. However, as indicated by the $\Delta\%$ values, the transformations alone are insufficient to achieve the same results as when they are combined.

\begin{figure*}
    \centering
    % \begin{subfigure}[b]{0.4\textwidth}
    %     \includegraphics[width=2in, trim = 0cm 0.5cm 0cm 0cm]{orig.pdf}
    %     \caption{\fq}%{CFG of the function \fs}
    %     \label{img:fq_example}
    % \end{subfigure}
    % \begin{subfigure}[b]{0.4\textwidth}
    %     \includegraphics[width=2in, trim = 0cm 0.5cm 0cm 0cm]{GEMINI_adv.pdf}
    %     \caption{\fadv against \gemini}%{CFG of the function \fadv}
    %     \label{img:fadv_gemini_example}
    % \end{subfigure}
    % \hfill
    % \newline
    % \begin{subfigure}[b]{0.4\textwidth}
    %     \includegraphics[width=2in, trim = 0cm 0.5cm 0cm 0cm]{SAFE_adv.pdf}
    %     \caption{\fadv against \safe}
    %     \label{img:fadv_safe_example}
    % \end{subfigure}
    % \begin{subfigure}[b]{0.4\textwidth}
    %     \includegraphics[width=2in, trim = 0cm 0.5cm 0cm 0cm]{JTRANS_adv.pdf}
    %     \caption{\fadv against \jtrans}
    %     \label{img:fadv_jtrans_example}
    % \end{subfigure}
    \includegraphics[width=0.90\linewidth]{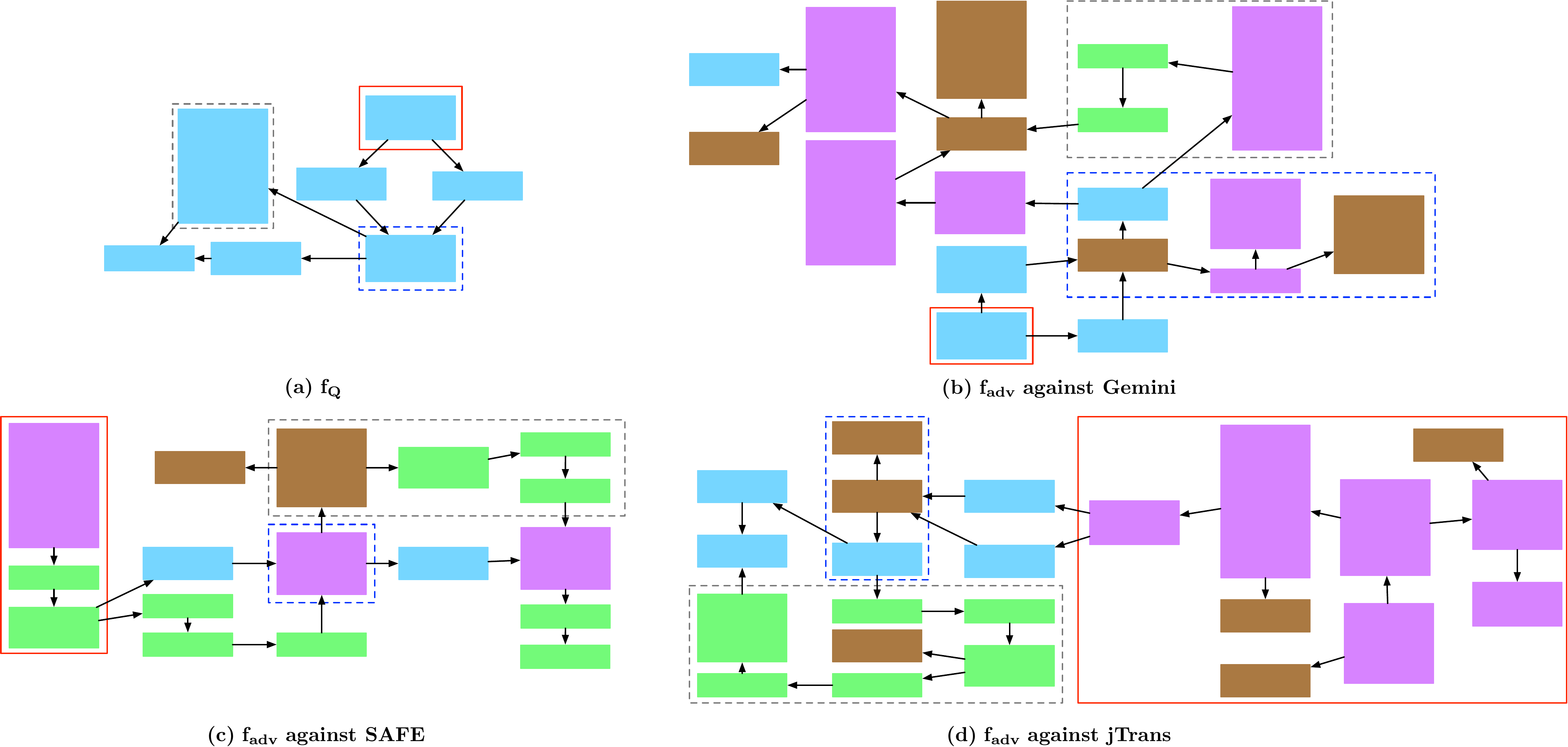}
    \caption{CFGs of the three binary functions in case of \textbf{untargeted} attack with $\lambda=0$ against \gemini, \safe, and \jtrans. \textbf{(a)} shows the CFG of \fq, while \textbf{(b)}, \textbf{(c)}, and \textbf{(d)} show the CFGs of the adversarial \fadv targeting \gemini, \safe, and \jtrans respectively. The \color{red}\textbf{red} \color{black} rectangle marks the function's entry point;
    \color{lightpurple}\textbf{purple} \color{black}, \color{mygreen}\textbf{green} \color{black}, and \color{brown}\textbf{brown} \color{black} blocks indicate the use of \textbf{SA}, \textbf{NS}, and \textbf{DBA}, respectively. Dotted rectangles highlight how instructions from a block in \fq are distributed across multiple blocks in \fadv after the attack.}
    \label{img:RQ4}
\end{figure*}

\subsubsection{Efficiency Analysis}\label{sec:eff}
We analyze our attack’s efficiency by comparing the average time needed to generate adversarial examples using different transformations, both individually and combined.

We observed that the execution time remains consistent between targeted and untargeted attacks since both run for a fixed number of iterations. Therefore, we focus only on the untargeted case. The main time factors in our attack are identifying perturbation positions, the number of candidate adversarial examples tested per iteration, and the overhead from binary function similarity calculations.

When using all transformations, the average execution time on \gemini, \safe, and \jtrans is $25.55 \pm 2.67$ minutes, indicating comparable generation time across models, with the procedure being more efficient against \gemini (taking 21.82 minutes in average) and least efficient against \safe (taking 27.91 minutes in average). Specifically, one iteration takes on average $50.12 \pm 6.03$ seconds.
 
When applied in isolation, transformations exhibit varying execution times. \textbf{IR} is the most efficient, taking $1.77 \pm 0.39$ minutes on average overall and $3.15 \pm 1.42$ seconds per iteration, while \textbf{DBA} is the least efficient, taking $17.17 \pm 4.65$ minutes overall and $34.07 \pm 9.80$ seconds per iteration.

\subsubsection{Qualitative Analysis}

Figure~\ref{img:RQ4} presents a comparison between the CFGs of the original query function \fq (shown in Figure~\ref{img:RQ4}(a)) and its adversarial versions generated after an untargeted attack with $\lambda = 0$ on three models: \gemini (Figure~\ref{img:RQ4}(b)), \safe (Figure~\ref{img:RQ4}(c)), and \jtrans (Figure~\ref{img:RQ4}(d)).

The sequence of transformations applied to generate the adversarial example in Figure~\ref{img:RQ4}(b) demonstrates that \gemini considers both the CFG topology and the individual instructions, as noted in the previous Section. A manual analysis of the adversarial example shows that the greedy optimizer typically first alters the topology with a combination of \textbf{DBA} and \textbf{NS} transformations, followed by adding new instructions using \textbf{SA}.

The adversarial example in Figure~\ref{img:RQ4}(c) shows that, when attacking \safe, most modifications are focused around the prologue of \fq. Specifically, the function's entry point is modified twice: first, by adding new instructions through the \textbf{SA} transformation, and then by splitting the final portion of the block with \textbf{NS}. Interestingly, one of the neighbors of the entry point is also modified using \textbf{NS}. This aligns with what has been observed in ~\cite{DBLP:conf/icse/WongWLW24}, which shows that \safe usually focuses its analysis on functions' prologue.

Figure~\ref{img:RQ4}(d) shows the adversarial function \fadv generated from \fq when targeting \jtrans. As noted in the previous section, the distribution of the different transformations is relatively balanced. Like \safe, the attack strategy focuses on the entry point, applying several transformations that add both new nodes and instructions.

The analysis of single nodes reveals how the greedy procedure tailors its modifications based on the target model. For the node shown in gray, when targeting \gemini, the modification involves adding a strand and splitting the final part using \textbf{NS}. In contrast, for \safe, the original instructions are spread across five new nodes, including a dead branch with additional instructions, created through a combination of \textbf{NS} and \textbf{DBA}. When attacking \jtrans, the approach resembles the one used for \safe, but applies \textbf{NS} more frequently. This last observation is expected because, as outlined in the previous section, \jtrans indirectly accounts for the CFG of a function by modeling \textit{jump} instructions differently than other types of instructions. The node in blue remains mostly unchanged when targeting \safe but undergoes similar modifications when targeting both \gemini and \jtrans.

\begin{mybox}
\textbf{Key takeaway.} The qualitative analysis confirms the findings from the distribution of applied transformations. Moreover, it uncovers hidden aspects of the target models, such as the tendency of certain models to concentrate on the prologue of functions.
\end{mybox}

\subsection{Non-ML Approaches}
In this section, we assess the robustness of non-ML methods for comparing binary functions. For this evaluation, we consider:

\begin{itemize}
    \item \textbf{GSIZE}. A simple approach that compares two functions based on the number of basic blocks.
    \item \textbf{GEDIT}. A simple approach that compares the CFGs of two functions using an approximated labeled edit distance measure~\cite{DBLP:journals/prl/FischerRB17} (i.e., the number of changes in terms of nodes and edges edits to transform a source CFG into a target one).
    \item \textbf{Catalog1}\footnote{\url{https://www.xorpd.net/pages/fcatalog.html}}. This approach uses fuzzy hashing, directly leveraging raw binary information. Specifically, it applies MinHash~\cite{broder1997resemblance} to encode groups of four consecutive function bytes into a fixed-size signature, on which similarity is computed using Jaccard.
\end{itemize}

We excluded other approaches, such as PSS~\cite{DBLP:conf/sigsoft/BenoitMB23} and BinDiff~\cite{dullien2005graph}, which operate at the program level rather than the function level, as they fall outside the scope of our threat model.

\begin{table}[h]
\centering
\caption{Untargeted and Targeted attacks against non-ML approaches, considering $K=10$, $|\mPool|=128$, and $\lambda = 0$.}
\resizebox{0.5\textwidth}{!}{%
\begin{tabular}{|l|c|r|r|r|r|r|r|r|r|}
    \cmidrule[\heavyrulewidth]{3-10}
    \multicolumn{1}{l}{\multirow{2}{*}{}} & \multicolumn{1}{l|}{\multirow{1}{*}{}} & \multicolumn{4}{c|}{\textbf{Untargeted}} & \multicolumn{4}{c|}{\textbf{Targeted}} \\
    \cmidrule{3-10}
    \multicolumn{1}{l}{} & \multicolumn{1}{l|}{} & \multicolumn{1}{c|}{GSIZE} & \multicolumn{1}{c|}{GEDIT} & \multicolumn{1}{c|}{\catalog} & \multicolumn{1}{c|}{AVG} & \multicolumn{1}{c|}{GSIZE} & \multicolumn{1}{c|}{GEDIT} & \multicolumn{1}{c|}{\catalog} & \multicolumn{1}{c|}{AVG} \\
    \toprule
    \toprule
    & \wasr & 82.77 & 81.59 & 40.02 & 68.13 & 27.17 & 18.65 & 42.85 & 29.56 \\
    & \crecall & 0.58 & 0.49 & 0.69 & 0.59 & $\blacksquare$ & $\blacksquare$ & $\blacksquare$ & $\blacksquare$ \\
    & \arecall & 0.17 & 0.18 & 0.60 & 0.32 & $\blacksquare$ & $\blacksquare$ & $\blacksquare$ & $\blacksquare$ \\
    \toprule
    \toprule
    \multirow{4}{*}{\textsf{\textbf{@1}}} & \clean & 79.92 & 84.85 & 64.79 & 76.52 & 18.05 & 26.41 & 46.22 & 30.23 \\
    & \asr & 91.41 & 87.96 & 53.52 & 77.63 & 47.16 & 25.80 & 73.71 & 48.89 \\
    & \modi & 108.08 & 4.26 & 94.87 & 69.07 & 35.91 & 11.68 & 55.64 & 34.41 \\
    & \modn & 40.60 & 2.0 & 5.60 & 16.07 & 14.16 & 3.23 & 1.94 & 6.44 \\
    \midrule
    \multirow{4}{*}{\textsf{\textbf{@2}}} & \clean & 62.34 & 77.83 & 44.57 & 61.58 & 10.07 & 17.47 & 22.41 & 16.65 \\
    & \asr & 87.31 & 86.66 & 50.0 & 74.66 & 33.40 & 18.98 & 52.29 & 34.89 \\
    & \modi & 111.79 & 4.25 & 95.82 & 70.62 & 44.65 & 15.67 & 63.32 & 41.21 \\
    & \modn & 41.93 & 0 & 5.55 & 15.83 & 17.50 & 4.30 & 2.24 & 8.01 \\
    \midrule
    \multirow{4}{*}{\textsf{\textbf{@3}}} & \clean & 25.17 & 43.33 & 15.29 & 27.93 & 3.19 & 13.55 & 12.05 & 9.60 \\
    & \asr & 80.22 & 81.95 & 36.82 & 66.33 & 17.45 & 15.36 & 28.49 & 20.43 \\
    & \modi & 117.77 & 4.27 & 102.15 & 74.73 & 61.18 & 19.19 & 67.59 & 49.32 \\
    & \modn & 43.94 & 2.0 & 5.15 & 17.03 & 23.90 & 5.26 & 2.57 & 10.58 \\
    \midrule
    \multirow{4}{*}{\textsf{\textbf{@4}}} & \clean & 0 & 0 & 0 & 0 & 1.40 & 12.15 & 8.86 & 7.47 \\
    & \asr & 72.13 & 69.81 & 19.72 & 53.89 & 10.67 & 14.46 & 16.93 & 14.02 \\
    & \modi & 127.10 & 4.27 & 109.67 & 80.35 & 75.16 & 20.39 & 75.81 & 57.12 \\
    & \modn  & 47.23 & 2.0 & 5.07 & 18.10 & 29.18 & 5.59 & 2.76 & 12.51 \\
    \bottomrule
\end{tabular}}

\vspace{-0.3cm}
\label{tab:nonML_K10}
\end{table}

Table~\ref{tab:nonML_K10} presents the results for the untargeted and targeted attacks against the three considered approaches. For the untargeted scenario, it is evident that graph-based approaches  are not robust against our strategy, with a \wasr of 82.77\% for GSIZE and 81.59\% for GEDIT. When moving to \catalog, this presents better robustness compared to the other considered approaches, with a \wasr of 40.02\%. In the targeted scenario, all the considered approaches exhibit greater robustness to our attack compared to DNN-based solutions. Specifically, the average \wasr against non-ML methods is 29.56\% (compared to 57.06\% for DNN-based solutions), with \catalog being the least robust (42.85\%) and GEDIT the most robust (18.65\%).

The previous results highlight the better robustness of non-ML approaches compared to DNN-based ones against our attack. However, it is worth noting that non-ML solutions present lower \crecall values, showing poor performance on clean data.

The poor attack performance in the targeted case against GSIZE and GEDIT is due to the fact that our transformations cannot remove nodes from the CFG. As a result, when the target function has fewer nodes than the query, the attack fails. This also highlights a significant limitation of these similarity methods, as they struggle to assign high similarity scores to functions that are semantically similar but topologically different.

Regarding \catalog, its robustness in the targeted case is comparable to that of DNN-based systems; however, it shows remarkable robustness in the untargeted case. Probably the set of our transformations is unable to effectively modify the representation used by \catalog. The results $@1$, where \cleanone is higher than \asrone, highlight that, for certain queries \fq, at least one of the variants initially ranked outside the top-$K$ is ranked in the top-$K$ of \fadv. This implies that \catalog fails to assign high similarity scores to semantically similar functions and that our attack indirectly improves the performance of the system on a limited number of clean data.

%% file: related_work.tex
% !TEX root =  main.tex

\section{Related Works}\label{sec:related}

In this section, we first discuss techniques for generating adversarial examples against image classifiers and NLP models; then, we move to approaches targeting models for source code analysis. Finally, we discuss attacks against malware detectors and models for binary analysis.

\subsection{Attacking Image Classifiers and NLP Models}

Adversarial attacks were first introduced against models for image classification, with early works~\cite{DBLP:conf/pkdd/BiggioCMNSLGR13, DBLP:journals/corr/SzegedyZSBEGF13, goodfellow2014explaining, carlini2017towards}  providing white-box, gradient-based methods that add minimal perturbations to fool models with high confidence. %whose goal is to insert into a clean instance the smallest possible perturbation that can mislead the target model. They essentially rely on gradient-driven approaches to create adversarial examples with maximum confidence. 
Chen et al.~\cite{DBLP:conf/sp/ChenJW20} propose various black-box decision-based attacks against image classifiers involving the estimation of gradient direction.

Jia et al.~\cite{DBLP:conf/emnlp/JiaL17} attack reading comprehension models by introducing sentences that can deceive the target models while maintaining the original semantics of the paragraph. More recently, the solutions in~\cite{DBLP:conf/emnlp/LiMGXQ20, DBLP:conf/naacl/LiZPCBSD21} proposed to attack NLP models by finding replacements of words composing the input sequence, using BERT-based strategies.

\subsection{Attacking Models for Source Code Analysis}

Methods for attacking models for source code analysis are mainly based on applying semantics-preserving perturbations at the source code level, thus having limited applicability in the binary function similarity context.

Yefet et al.~\cite{yefet2020adversarial} propose a white-box approach that, using a gradient-driven method, iteratively changes the names of variables defined within a function in all their occurrences, until a misclassification occurs. Differently, Zhang et al.~\cite{zhang2023challenging} target code clone detectors using semantics-preserving transformations, combined using common optimization heuristics, alongside a reinforcement learning-based approach for searching clones that could evade the detection.

\subsection{Attacking Models for Binary Code Analysis}
The solution proposed by Pierazzi et al.~\cite{pierazzi2020intriguing} targets Android malware classifiers and is based on software transplantation. Here, benign snippets of code that can trigger the classifier features are injected into the malware sample to cause a misclassification using a gradient-guided approach. Moreover, the snippets are injected into portions of code that are never executed, to guarantee the preservation of the semantics.

Lucas et al.~\cite{lucas2021malware} attack malware classifiers based on raw bytes. Their solution is based on combining different semantics-preserving perturbations both in a black-box and a white-box context. The proposed transformations are a subset of ours, as they include \textbf{IR} and \textbf{NS}; however, while these transformations show clear effectiveness when targeting malware classifiers (and also commercial solutions), they may show poor performance when targeting binary function similarity models; indeed, it is evident from our results that a crucial point in attacking binary similarity models is the need of inserting new instructions into the function being modified.

FuncFooler~\cite{jia2022funcfooler} is a recent unpublished work targeting binary function similarity models in the context of function search. It consists of an instruction-insertion strategy to modify a binary function at a set of fixed locations determined in advance; to guarantee semantics preservation, possible side effects are corrected a posteriori. The set of possible instructions is computed considering the instructions of the functions in the pool. Differently from our approach, FuncFooler explores only one class of transformations (which can be considered a subset of \textbf{SA} where a single instruction is inserted at each step), without altering the topology of the CFG. Furthermore, it only studies untargeted attacks.

Capozzi et al.~\cite{capozzi2023adversarial} is a recent unpublished work proposing two solutions targeting a subset of the binary function similarity models we considered. Their black-box approach consists of a greedy solution, feasible both in a targeted and untargeted context, that iteratively inserts new instructions into dead branches, whose locations are fixed in advance. Differently from FuncFooler~\cite{jia2022funcfooler}, the set of possible instructions is dynamically updated using a heuristic based on instruction embeddings. The proposed white-box attack substitutes the aforementioned heuristic with a gradient-guided instruction insertion strategy. We highlight that~\cite{capozzi2023adversarial} relies on a set of transformations that is strictly a subset of ours (as~\cite{capozzi2023adversarial} uses only {\bf DBA} with a single instruction added) and it tests its approach only against three models (namely, \gemini, \safe, and \gmn).

PELICAN~\cite{DBLP:conf/uss/0002TSAXLYW023} is a novel white-box attack that leverages natural backdoors of attacked models to identify instructions that, once inserted into functions, can induce misclassification. This attack has been tested against models for different binary analysis tasks, including function naming, compiler provenance, and binary function similarity. In the context of the latter, the proposed methodology has only been tested against three models---specifically, \gemini, \safe, and \trex.

We emphasize that the last two solutions differ significantly from our work. Firstly, they do not address the function search task. Specifically, Capozzi et al.~\cite{capozzi2023adversarial} target the similarity function implemented by the target model, whereas PELICAN~\cite{DBLP:conf/uss/0002TSAXLYW023} focuses on attacking the loss function implemented by the target model. Another key difference lies in the use of variants of the query function during the optimization process, which is not the case in either of the other two approaches.

\section{Discussion}
We now discuss the practical impacts of our study and the limitations of our evaluation setting.

\subsection{Practical Impacts}
As outlined in Section~\ref{sec:introduction}, binary function similarity systems play a crucial role in various security-sensitive scenarios, including vulnerability detection, plagiarism identification, and malware analysis. These systems help automate the process of comparing binary functions, making it easier to identify code reuse, detect security flaws, and uncover malicious behaviors. In practical scenarios, such systems are integrated into tools used by reverse engineers. Notable examples include plugins such as YARASAFE\footnote{\url{https://github.com/lucamassarelli/yarasafe}} and BinaryAI\footnote{\url{https://github.com/binaryai/plugins}}. These plugins facilitate the use of traditional reverse engineering tools by providing automated capabilities that can reduce manual effort.

Our threat model represents a practical scenario where a remotely deployed binary function similarity system operates as a black-box model, providing only similarity scores. While this represents a worst-case assumption for the attacker, our findings reveal that these systems remain vulnerable to our attack, which is relatively simple to implement in practical contexts. This applies to both targeted (e.g., disguising malicious code as benign) and untargeted (e.g., hiding vulnerable or plagiarized functions) attacks, posing serious threats in real-world scenarios, even for models explicitly designed to handle obfuscated functions, such as \binfinder~\cite{10.1145/3579856.3582818} and \trex~\cite{DBLP:journals/tse/PeiXYJR23}.

\subsection{Limitations}
We examine the limitations of our work, focusing on the dataset and transformations used. Our dataset is smaller than benchmarks like BinaryCorp~\cite{jTrans-ISSTA22}, BinKit~\cite{kim2022revisiting}, and those used by Marcelli et al.~\cite{marcelli2022machine}. However, these benchmarks are typically used to evaluate the performance of binary function similarity systems on clean data. Due to the computational cost of generating adversarial examples (see Section~\ref{sec:eff}), using such large datasets is impractical in our scenario. However, the number of open-source projects used to generate our codebase aligns with standard practices in the field, as prior studies~\cite{DBLP:conf/icml/LiGDVK19, massarelli2021function, DBLP:journals/tse/PeiXYJR23, 10.1145/3579856.3582818} typically extract functions from 1 to 10 projects.

%Compared to established benchmarks in the field, such as BinaryCorp~\cite{jTrans-ISSTA22}, BinKit~\cite{kim2022revisiting}, and the datasets used by Marcelli et al.~\cite{marcelli2022machine}, our dataset is of a more limited size. However, the number of open-source projects we considered for generating our codebase of functions aligns with standard practices in the field, where most of the related works~\cite{DBLP:conf/icml/LiGDVK19, massarelli2021function, DBLP:journals/tse/PeiXYJR23, 10.1145/3579856.3582818} typically use between 1 and 10 projects to extract binary functions.

%The variety of a binary dataset depends on the number of architectures and compilers considered. Regarding architectures, 
% As most binary function similarity systems are trained exclusively on ELF \texttt{amd64} functions compiled from C code, we limited our evaluation to this setting, excluding other ISAs and source code languages. However, our attack is architecture-agnostic and can be extended to other ISAs by adapting the semantics-preserving transformations. We expect our findings to generalize, as the attack does not rely on ISA-specific traits. Furthermore, as the considered models have not been trained on such binaries, then their performance on clean data may degrade, potentially increasing the ASR.
As most binary function similarity systems are trained on ELF \texttt{amd64} functions compiled from C code, we limited our evaluation to this setting. However, our attack is architecture-agnostic and can be extended to other ISAs by adapting the transformations. We expect our findings to generalize, as the attack does not depend on ISA-specific traits. Furthermore, variations in source code languages may alter assembly representations, and if models are not trained on such binaries, their performance may degrade, potentially increasing the ASR.
With respect to compilers, binary function similarity systems are typically trained considering multiple compilers as well as different versions of the same compiler. While our dataset accounts for the first aspect, we did not explore the latter. However, \cite{marcelli2022machine} observed a slight performance drop on clean data when comparing functions compiled with different versions of the same compiler, suggesting that the \asr would likely increase in such scenarios.

Finally, our set of transformations may have little to no effect against symbolic execution-based methods. However, these methods are often impractical due to their inefficiency; indeed, comparing binary functions using symbolic execution may lead to path explosion, making it impractical in real-world scenarios.

\section{Conclusions and Future Works}

In this paper, we presented the first large-scale analysis of the robustness of binary function similarity models against adversarial attacks highlighting the need for a trade-off between performance and robustness.

We demonstrated that a simple greedy strategy, when enriched with a wide set of transformations, can mount untargeted attacks with very high success rates on all considered models, particularly those showing top performance on clean data. Conversely, models that initially perform poorly seem to be more resistant to adversarial examples. On the targeted front, our attacks performed slightly worse, but they were still successful in more than half of the instances considered.

We investigated several additional aspects. First, we showed that adversarial examples transfer across models, with a significantly higher success in the untargeted case rather than the targeted. Secondly, we demonstrated that the set of transformations we considered was effective in modifying most of the key features considered by the target models, with two transformations making a particularly strong contribution to the success of the attack. Finally, manual analysis of adversarial examples uncovered hidden behaviors in the models, revealing that they focus their analysis on specific portions of the functions.

Our research opens several new research avenues, particularly in the context of defense strategies. Rather than focusing solely on adversarial training--- which may enhance model robustness against our attack but does not guarantee protection against zero-day threats--- we argue that greater emphasis should be placed on proposing inherently robust function representation methods.

%% file: appendix.tex
% !TEX root = main.tex

\section{}\label{sec:app_C}
Algorithm~\ref{pseudo} presents our attack procedure. The first adversarial example \fadv is the query function \fq itself (line~\ref{alg:initadv}). We then initialize the set of candidate strands that can be inserted into the adversarial function either using \textbf{DBA} or \textbf{SA} (line~\ref{alg:strandinit}). Then, during the iterative procedure, we first identify possible positions to perturb (line~\ref{alg:posupdate}) and then enumerate all the possible transformations that can be applied in the identified positions. Specifically, we apply a transformation $tr$ at the position $pos$, for every pair $\langle tr, pos \rangle \in TR \times POS$ (lines~\ref{alg:eninit} -~\ref{alg:enend}). We then proceed to evaluate the objective function defined in Equation~\ref{eq:obj}, considering the set of candidates $CAND$ and the set of target variants $V$ (line~\ref{alg:objeval}). Finally, we select the new adversarial example \fadv according to the value of $\varepsilon$ (lines~\ref{alg:epsstart} -~\ref{alg:epsend}) and update the set of candidate strands (line~\ref{alg:strandupd}). The final adversarial example \fadv (line~\ref{alg:finaladv}) is, among all \fadv generated at the end of each iteration, the one that produced the highest value for the objective function.

\begin{algorithm}[h]
\begin{small}
\footnotesize
\caption{Greedy Optimization Strategy}
\label{pseudo}
\textbf{Input:}
\begin{itemize}
    \item Query function \fq
    % \item Pool of function \pool
    \item Set of target variants $V$
\end{itemize}
\textbf{Output:} Adversarial example \fadv\\
\textbf{Definitions:}
\begin{itemize}
    \item Maximum number of iterations $\Delta$
    \item Set of semantics-preserving transformations $TR$
    \item randomStrands(): Initialize the set $STRANDS$ with random strands.
    \item $\mathrm{ir}(\mFadv, pos)$. Apply the IR transformation to \fadv at location $pos$. Return a new candidate adversarial example.
    \item $\mathrm{ns}(\mFadv, pos)$. Apply the NS transformation to \fadv at location $pos$. Return a new candidate adversarial example.
    \item $\mathrm{dba}(\mFadv, pos)$. Apply the DBA transformation to \fadv at location $pos$. Return a list of $|STRANDS|$ candidate adversarial examples, where each candidate consists of adding in a dead branch at position $pos$ within \fadv a strand from $STRANDS$.
    \item $\mathrm{sa}(\mFadv, pos)$. Apply the SA transformation to \fadv at location $pos$. Return a list of $|STRANDS|$ candidate adversarial examples, where each candidate consists of adding at position $pos$ within \fadv a strand from $STRANDS$.
    \item $\mathrm{evaluate}(cands, V)$. Evaluate the objective function in Equation~\ref{eq:obj} considering the possible candidates $cands$ and the set of target variants $V$.
    \item $\mathrm{best}(advs)$. Given a set of adversarial examples, return the one that maximizes the value of Equation~\ref{eq:obj}.
\end{itemize}
\begin{algorithmic}[1]
    \State $\mFadv \leftarrow \mFq$\label{alg:initadv}
    \State $iter \leftarrow 0$
    \State $STRANDS \leftarrow \mathrm{randomStrands}()$\label{alg:strandinit}
    \State $advs, POS \leftarrow [ \, ],  [ \, ]$
    \While{$iter < \Delta$}
        \State $cands \leftarrow [ \, ]$
        \State $POS.\mathrm{update}()$\label{alg:posupdate}
        \For{$\langle tr, pos \rangle \in TR \times POS$}\label{alg:eninit}
            \If{$tr ==$ `IR'}
                \State $cands.\mathrm{extends}(\mathrm{ir}(\mFadv, pos))$
            \ElsIf{$tr ==$ `NS'}
                \State $cands.\mathrm{extends}(\mathrm{ns}(\mFadv, pos))$
            \ElsIf{$tr ==$ `DBA'}
                \State $cands.\mathrm{extends}(\mathrm{dba}(\mFadv, pos))$
            \Else
                \State $cands.\mathrm{extends}(\mathrm{sa}(\mFadv, pos))$
            \EndIf
        \EndFor\label{alg:enend}
        \State $objective\_values \leftarrow \mathrm{evaluate}(cands, V)$\label{alg:objeval}
        \State $prob \leftarrow uniform(0, 1)$\label{alg:epsstart}
        \If{$prob < \varepsilon$}
            \State $\mFadv \leftarrow \mathrm{selectGreedy}(objective\_values)$
        \Else
            \State $\mFadv \leftarrow \mathrm{selectRandom}(objective\_values)$
        \EndIf\label{alg:epsend}
        \State $advs.\mathrm{extends}(\mFadv)$
        \State $STRANDS.\mathrm{update}(objective\_values)$\label{alg:strandupd}
        \State $iter \leftarrow iter + 1$
    \EndWhile
    \State \Return $\mathrm{best}(advs)$\label{alg:finaladv}
\end{algorithmic}

\end{small}
\end{algorithm}
\newpage

\section{}\label{sec:app_A}

Below, we present the results of the robustness analysis for the evaluated models, considering various values of $K$, $\mPool$, and $\lambda$. Specifically, we set $K \in \{10, 100\}$ for untargeted attacks and $K \in \{5, 10\}$ for targeted attacks. Additionally, we consider pools with sizes $|\mPool| \in \{32, 128, 512, 1000\}$ and $\lambda \in \{0, 0.01, 0.3\}$.

We report the results for the untargeted case in Tables~\ref{tab:Untar_32}, \ref{tab:Untar_128}, \ref{tab:Untar_512}, \ref{tab:Untar_1000} and \ref{tab:Untar_001}, while the ones for the targeted scenario in Tables~\ref{tab:Tar_32}, \ref{tab:Tar_128}, \ref{tab:Tar_512}, and \ref{tab:Tar_1000}.

\begin{table*}[]
\centering
\footnotesize
\caption{Untargeted attack at $K=10$ when considering a pool of size 32 with $\lambda \in \{0, 0.3\}$. In column AVG we report the average of the measures across all models.}
\resizebox{\textwidth}{!}{%
\begin{tabular}{|c|c|c|rr|rr|rr|rr|rr|rr|rr|rr|rr|} 
        \cmidrule[\heavyrulewidth]{4-19}
        \multicolumn{1}{l}{} & \multicolumn{1}{l}{\multirow{3}{*}{}} & \multicolumn{1}{l|}{\multirow{2}{*}{}} & \multicolumn{16}{c|}{\textbf{Models}} & \multicolumn{2}{c}{\multirow{1}{*}{}} \\
        \cline{4-21}
        \multicolumn{1}{l}{} & \multicolumn{1}{l}{} & \multicolumn{1}{l|}{} & \multicolumn{2}{c|}{\gemini} & \multicolumn{2}{c|}{\gmn} & \multicolumn{2}{c|}{\zeek} & \multicolumn{2}{c|}{\binfinder} & \multicolumn{2}{c|}{\safe} & \multicolumn{2}{c|}{\jtrans} & \multicolumn{2}{c|}{\trex} & \multicolumn{2}{c|}{\palmtree} & \multicolumn{2}{c|}{AVG} \\
        \cline{3-21}
        \multicolumn{1}{l}{} & \multicolumn{1}{l|}{} & \bm{$\lambda$} & \textbf{0.0} & \textbf{0.3} & \textbf{0.0} & \textbf{0.3} & \textbf{0.0} & \textbf{0.3} & \textbf{0.0} & \textbf{0.3} & \textbf{0.0} & \textbf{0.3} & \textbf{0.0} & \textbf{0.3} & \textbf{0.0} & \textbf{0.3} & \textbf{0.0} & \textbf{0.3} & \textbf{0.0} & \textbf{0.3} \\
        \toprule
        \toprule
        &  & \crecall & 0.87 & 0.87 & 0.98 & 0.98 & 0.92 & 0.92 & 0.96 & 0.96 & 0.99 & 0.99 & 0.94 & 0.94 & 0.99 & 0.99 & 0.98 & 0.98 & 0.95 & 0.95 \\
        &  & \arecall & 0.15 & 0.85 & 0.12 & 0.90 & 0.33 & 0.82 & 0.12 & 0.96 & 0.07 & 0.97 & 0.31 & 0.90 & 0.14 & 0.98 & 0.11 & 0.97 & 0.17 & 0.92 \\
        &  & \wasr & 85.07 & 15.47 & 88.05 & 9.60 & 66.75 & 18.14 & 88.28 & 3.77 & 93.15 & 3.32 & 69.15 & 9.98 & 86.38 & 2.23 & 88.78 & 3.13 & 83.20 & 8.20 \\
        \cline{2-21}
        & \multirow{4}{*}{} & \clean & 34.67 & 34.67 & 4.70 & 4.70 & 22.26 & 22.26 & 9.80 & 9.80 & 5.0 & 5.0 & 21.0 & 21.0 & 2.58 & 2.58 & 5.81 & 5.81 & 13.23 & 13.23 \\
        &  & \asr & 91.18 & 37.70 & 97.80 & 18.50 & 82.65 & 39.68 & 91.80 & 10.10 & 96.90 & 8.75 & 94.90 & 28.40 & 93.04 & 5.80 & 93.99 & 9.22 & 92.78 & 19.77 \\
\bottomrule
    \end{tabular}}
\vspace{-0.3cm}
\label{tab:Untar_32}
\end{table*}

\begin{table*}[]
\centering
\footnotesize
\caption{Untargeted attack at $K=10$ and $K=100$ when considering a pool of size 128 with $\lambda \in \{0, 0.3\}$. In column AVG we report the average of the measures across all models.}
\resizebox{\textwidth}{!}{%
    \begin{tabular}{|c|c|c|rr|rr|rr|rr|rr|rr|rr|rr|rr|} 
        \cmidrule[\heavyrulewidth]{4-19}
        \multicolumn{1}{l}{} & \multicolumn{1}{l}{\multirow{3}{*}{}} & \multicolumn{1}{l|}{\multirow{2}{*}{}} & \multicolumn{16}{c|}{\textbf{Models}} & \multicolumn{2}{c}{\multirow{1}{*}{}} \\
        \cline{4-21}
        \multicolumn{1}{l}{} & \multicolumn{1}{l}{} & \multicolumn{1}{l|}{} & \multicolumn{2}{c|}{\gemini} & \multicolumn{2}{c|}{\gmn} & \multicolumn{2}{c|}{\zeek} & \multicolumn{2}{c|}{\binfinder} & \multicolumn{2}{c|}{\safe} & \multicolumn{2}{c|}{\jtrans} & \multicolumn{2}{c|}{\trex} & \multicolumn{2}{c|}{\palmtree} & \multicolumn{2}{c|}{AVG} \\
        \cline{3-21}
        \multicolumn{1}{l}{} & \multicolumn{1}{l|}{} & \bm{$\lambda$} & \textbf{0.0} & \textbf{0.3} & \textbf{0.0} & \textbf{0.3} & \textbf{0.0} & \textbf{0.3} & \textbf{0.0} & \textbf{0.3} & \textbf{0.0} & \textbf{0.3} & \textbf{0.0} & \textbf{0.3} & \textbf{0.0} & \textbf{0.3} & \textbf{0.0} & \textbf{0.3} & \textbf{0.0} & \textbf{0.3} \\
        \toprule
        \toprule
        &  & \crecall & 0.71 & 0.71 & 0.94 & 0.94 & 0.73 & 0.73 & 0.91 & 0.91 & 0.94 & 0.94 & 0.81 & 0.81 & 0.95 & 0.95 & 0.90 & 0.90 & 0.86 & 0.86 \\
        &  & \arecall & 0.07 & 0.65 & 0.03 & 0.80 & 0.05 & 0.61 & 0.06 & 0.9 & 0.02 & 0.86 & 0.04 & 0.75 & 0.04 & 0.92 & 0.03 & 0.86 & 0.04 & 0.79 \\
        &  & \wasr & 92.99 & 35.05 & 97.30 & 20.47 & 95.23 & 39.30 & 93.80 & 10.27 & 98.30 & 13.51 & 96.0 & 25.20 & 96.17 & 8.30 & 96.69 & 13.80 & 95.81 & 20.74 \\
        \cline{2-21}
        &  & \clean & 63.63 & 63.63 & 16.10 & 16.10 & 61.76 & 61.76 & 24.40 & 24.40 & 19.40 & 19.40 & 51.10 & 51.10 & 12.82 & 12.82 & 27.66 & 27.66 & 34.61 & 34.61 \\
        &  & \asr & 96.69 & 67.50 & 99.30 & 35.10 & 98.06 & 68.52 & 95.80 & 25.80 & 99.20 & 31.19 & 99.40 & 55.80 & 97.61 & 19.40 & 98.0 & 33.97 & 98.01 & 42.16 \\
\bottomrule
    \end{tabular}}
\vspace{-0.3cm}
\label{tab:Untar_128}
\end{table*}

\begin{table*}[]
\centering
\footnotesize
\caption{Untargeted attack at $K=10$ adn $K=100$ when considering a pool of size 512 with $\lambda \in \{0, 0.3\}$. In column AVG we report the average of the measures across all models.}
\resizebox{\textwidth}{!}{%
    \begin{tabular}{|c|c|c|rr|rr|rr|rr|rr|rr|rr|rr|rr|} 
        \cmidrule[\heavyrulewidth]{4-19}
        \multicolumn{1}{l}{} & \multicolumn{1}{l}{\multirow{3}{*}{}} & \multicolumn{1}{l|}{\multirow{2}{*}{}} & \multicolumn{16}{c|}{\textbf{Models}} & \multicolumn{2}{c}{\multirow{1}{*}{}} \\
        \cline{4-21}
        \multicolumn{1}{l}{} & \multicolumn{1}{l}{} & \multicolumn{1}{l|}{} & \multicolumn{2}{c|}{\gemini} & \multicolumn{2}{c|}{\gmn} & \multicolumn{2}{c|}{\zeek} & \multicolumn{2}{c|}{\binfinder} & \multicolumn{2}{c|}{\safe} & \multicolumn{2}{c|}{\jtrans} & \multicolumn{2}{c|}{\trex} & \multicolumn{2}{c|}{\palmtree} & \multicolumn{2}{c|}{AVG} \\
        \cline{3-21}
        \multicolumn{1}{l}{} & \multicolumn{1}{l|}{} & \bm{$\lambda$} & \textbf{0.0} & \textbf{0.3} & \textbf{0.0} & \textbf{0.3} & \textbf{0.0} & \textbf{0.3} & \textbf{0.0} & \textbf{0.3} & \textbf{0.0} & \textbf{0.3} & \textbf{0.0} & \textbf{0.3} & \textbf{0.0} & \textbf{0.3} & \textbf{0.0} & \textbf{0.3} & \textbf{0.0} & \textbf{0.3} \\
        \toprule
        \toprule
        \multirow{19}{*}{} &  & \crecall & 0.56 & 0.56 & 0.86 & 0.86 & 0.51 & 0.51 & 0.83 & 0.83 & 0.84 & 0.84 & 0.67 & 0.67 & 0.89 & 0.89 & 0.79 & 0.79 & 0.74 & 0.74 \\
        &  & \arecall & 0.04 & 0.49 & 0.01 & 0.66 & 0.02 & 0.39 & 0.04 & 0.82 & 0.0 & 0.73 & 0.01 & 0.59 & 0.02 & 0.83 & 0.01 & 0.72 & 0.02 & 0.65 \\
        &  & \wasr & 96.27 & 51.45 & 99.0 & 33.65 & 98.12 & 60.78 & 96.33 & 18.0 & 99.58 & 26.86 & 99.15 & 41.38 & 98.43 & 17.05 & 99.37 & 28.33 & 98.28 & 34.69 \\
        \cline{2-21}
        & \multirow{4}{*}{} & \clean & 82.87 & 82.87 & 36.40 & 36.40 & 84.70 & 84.70 & 39.40 & 39.40 & 42.60 & 42.60 & 73.20 & 73.20 & 29.22 & 29.22 & 53.15 & 53.15 & 55.19 & 55.19 \\
        &  & \asr & 98.10 & 85.30 & 99.70 & 57.30 & 99.09 & 88.77 & 97.80 & 40.20 & 100.0 & 54.23 & 100.0 & 77.60 & 99.30 & 37.50 & 99.90 & 61.72 & 99.24 & 62.83 \\
\bottomrule
    \end{tabular}}
\vspace{-0.3cm}
\label{tab:Untar_512}
\end{table*}

\begin{table*}[]
\centering
\footnotesize
\caption{Untargeted attack at $K=10$ and $K=100$ when considering a pool of size 1000 with $\lambda \in \{0, 0.3\}$. In column AVG we report the average of the measures across all models.}
\resizebox{\textwidth}{!}{%
    \begin{tabular}{|c|c|c|rr|rr|rr|rr|rr|rr|rr|rr|rr|} 
        \cmidrule[\heavyrulewidth]{4-19}
        \multicolumn{1}{l}{} & \multicolumn{1}{l}{\multirow{3}{*}{}} & \multicolumn{1}{l|}{\multirow{2}{*}{}} & \multicolumn{16}{c|}{\textbf{Models}} & \multicolumn{2}{c}{\multirow{1}{*}{}} \\
        \cline{4-21}
        \multicolumn{1}{l}{} & \multicolumn{1}{l}{} & \multicolumn{1}{l|}{} & \multicolumn{2}{c|}{\gemini} & \multicolumn{2}{c|}{\gmn} & \multicolumn{2}{c|}{\zeek} & \multicolumn{2}{c|}{\binfinder} & \multicolumn{2}{c|}{\safe} & \multicolumn{2}{c|}{\jtrans} & \multicolumn{2}{c|}{\trex} & \multicolumn{2}{c|}{\palmtree} & \multicolumn{2}{c|}{AVG} \\
        \cline{3-21}
        \multicolumn{1}{l}{} & \multicolumn{1}{l|}{} & \bm{$\lambda$} & \textbf{0.0} & \textbf{0.3} & \textbf{0.0} & \textbf{0.3} & \textbf{0.0} & \textbf{0.3} & \textbf{0.0} & \textbf{0.3} & \textbf{0.0} & \textbf{0.3} & \textbf{0.0} & \textbf{0.3} & \textbf{0.0} & \textbf{0.3} & \textbf{0.0} & \textbf{0.3} & \textbf{0.0} & \textbf{0.3} \\
        \toprule
        \toprule
        &  & \crecall & 0.49 & 0.49 & 0.81 & 0.81 & 0.0 & 0.0 & 0.79 & 0.79 & 0.78 & 0.78 & 0.60 & 0.60 & 0.85 & 0.85 & 0.72 & 0.72 & 0.63 & 0.63 \\
        &  & \arecall & 0.03 & 0.42 & 0.01 & 0.59 & 0.0 & 0.0 & 0.03 & 0.78 & 0.0 & 0.65 & 0.01 & 0.52 & 0.01 & 0.78 & 0.0 & 0.65 & 0.01 & 0.55 \\
        &  & \wasr & 97.39 & 58.28 & 99.45 & 41.33 & 100.0 & 100.0 & 97.27 & 22.27 & 99.87 & 34.53 & 99.50 & 47.57 & 98.78 & 22.39 & 99.60 & 34.89 & 98.98 & 45.16 \\
        \cline{2-21}
        & \multirow{4}{*}{} & \clean & 88.76 & 88.76 & 47.42 & 47.42 & 100.0 & 100.0 & 46.29 & 46.29 & 55.41 & 55.41 & 80.66 & 80.66 & 37.82 & 37.82 & 64.76 & 64.76 & 65.14 & 65.14 \\
        &  & \asr & 98.80 & 90.30 & 99.80 & 67.30 & 100.0 & 100.0 & 98.60 & 47.49 & 100.0 & 67.04 & 100.0 & 83.37 & 99.40 & 47.90 & 100.0 & 71.21 & 99.57 & 71.83 \\
\bottomrule
    \end{tabular}}
\vspace{-0.3cm}
\label{tab:Untar_1000}
\end{table*}

\begin{table*}[]
\centering
\footnotesize
\caption{Untargeted attack at $K=10$ and $K=100$ when considering $\lambda=0.01$ and $|\mPool| \in \{32, 128, 512, 1000\}$. In column AVG we report the average of the measures across all models.}
\resizebox{\textwidth}{!}{%
    \begin{tabular}{|l|l|c|rrrr|rrrr|rrrr|rrrr|}
    \cmidrule[\heavyrulewidth]{4-15}
    \multicolumn{1}{l}{} & \multicolumn{1}{l}{} & \multicolumn{1}{l|}{} & \multicolumn{12}{c|}{\textbf{Models}} & \multicolumn{1}{c}{} & \multicolumn{1}{c}{} & \multicolumn{1}{c}{} & \multicolumn{1}{c}{} \\
    \cline{4-19}
    \multicolumn{1}{l}{} & \multicolumn{1}{l}{} & \multicolumn{1}{l|}{} & \multicolumn{4}{c|}{\gemini} & \multicolumn{4}{c|}{\safe} & \multicolumn{4}{c|}{\jtrans} & \multicolumn{4}{c|}{AVG} \\
    \cline{3-19}
    \multicolumn{1}{l}{} & \multicolumn{1}{l|}{} & \textbf{Pool Size} & 32 & 128 & 512 & 1000 & 32 & 128 & 512 & 1000 & 32 & 128 & 512 & 1000 & 32 & 128 & 512 & 1000 \\
    \toprule
    \toprule
    &  & \crecall & 0.90 & 0.71 & 0.55 & 0.48 & 0.98 & 0.93 & 0.84 & 0.78 & 0.93 & 0.81 & 0.67 & 0.61 & 0.94 & 0.82 & 0.69 & 0.62 \\
    &  & \arecall & 0.46 & 0.23 & 0.14 & 0.11 & 0.26 & 0.07 & 0.02 & 0.01 & 0.66 & 0.34 & 0.17 & 0.12 & 0.46 & 0.21 & 0.11 & 0.08 \\
    &  & \wasr & 54.47 & 76.58 & 86.24 & 89.22 & 74.0 & 92.74 & 98.35 & 98.98 & 34.16 & 65.84 & 82.61 & 87.50 & 54.21 & 78.39 & 89.07 & 91.90 \\
    \cline{2-19}
    & \multirow{4}{*}{} & \clean & 27.51 & 63.15 & 83.94 & 89.24 & 4.91 & 19.54 & 42.79 & 53.86 & 21.69 & 51.31 & 73.19 & 80.30 & 18.04 & 44.67 & 66.64 & 74.47 \\
    &  & \asr & 72.79 & 90.46 & 96.08 & 97.23 & 87.07 & 96.79 & 99.40 & 99.76 & 66.16 & 91.06 & 97.29 & 98.52 & 75.34 & 92.77 & 97.59 & 98.50 \\
\bottomrule
    \end{tabular}}
\vspace{-0.3cm}
\label{tab:Untar_001}
\end{table*}

\begin{table*}[]
\centering
\footnotesize
\caption{Targeted attack at $K=5$ and $K=10$ when considering a pool of size 32 with $\lambda \in \{0, 0.3\}$. In column AVG we report the average of the measures across all models.}
\resizebox{\textwidth}{!}{%
    \begin{tabular}{|c|c|c|rr|rr|rr|rr|rr|rr|rr|rr|rr|} 
        \cmidrule[\heavyrulewidth]{4-19}
        \multicolumn{1}{l}{} & \multicolumn{1}{l}{\multirow{3}{*}{}} & \multicolumn{1}{l|}{\multirow{2}{*}{}} & \multicolumn{16}{c|}{\textbf{Models}} & \multicolumn{2}{c}{\multirow{1}{*}{}} \\
        \cline{4-21}
        \multicolumn{1}{l}{} & \multicolumn{1}{l}{} & \multicolumn{1}{l|}{} & \multicolumn{2}{c|}{\gemini} & \multicolumn{2}{c|}{\gmn} & \multicolumn{2}{c|}{\zeek} & \multicolumn{2}{c|}{\binfinder} & \multicolumn{2}{c|}{\safe} & \multicolumn{2}{c|}{\jtrans} & \multicolumn{2}{c|}{\trex} & \multicolumn{2}{c|}{\palmtree} & \multicolumn{2}{c|}{AVG} \\
        \cline{3-21}
        \multicolumn{1}{l}{} & \multicolumn{1}{l|}{} & \bm{$\lambda$} & 0.0 & 0.3 & 0.0 & 0.3 & 0.0 & 0.3 & 0.0 & 0.3 & 0.0 & 0.3 & 0.0 & 0.3 & 0.0 & 0.3 & 0.0 & 0.3 & 0.0 & 0.3 \\
        \toprule
        \toprule
        \multicolumn{1}{|l|}{} & \multicolumn{1}{l|}{}  & \wasr & 58.05 & 14.44 & 68.10 & 15.20 & 38.14 & 13.36 & 92.11 & 14.58 & 74.17 & 17.72 & 56.05 & 22.65 & 62.78 & 15.24 & 73.25 & 15.37 & 65.33 & 16.07 \\
        \cline{2-21}
        &  & \clean & 30.10 & 30.10 & 31.0 & 31.0 & 29.29 & 29.29 & 25.25 & 25.25 & 27.10 & 27.10 & 43.60 & 43.60 & 22.30 & 22.30 & 29.16 & 29.16 & 29.72 & 29.72 \\
        &  & \asr & 85.70 & 32.70 & 85.70 & 33.60 & 59.69 & 29.66 & 97.8 & 26.15 & 90.40 & 34.74 & 94.0 & 54.80 & 80.70 & 28.31 & 90.38 & 31.59 & 85.55 & 33.94 \\
        & \multirow{-4}{*}{\textsf{\textbf{@1}}} & \modn & 13.64 & 0.92 & 18.68 & 0.96 & 13.72 & 1.05 & 33.50 & 0.07 & 15.57 & 0.98 & 11.50 & 1.04 & 11.40 & 0.90 & 18.01 & 0.85 & 17.0 & 0.85 \\
        \cline{2-21}
        &  & \clean & 16.80 & 16.80 & 15.0 & 15.0 & 12.65 & 12.65 & 16.23 & 16.23 & 17.10 & 17.10 & 20.0 & 20.0 & 14.60 & 14.60 & 16.33 & 16.33 & 16.09 & 16.09 \\
\bottomrule
    \end{tabular}}
\vspace{-0.3cm}
\label{tab:Tar_32}
\end{table*}

\begin{table*}[]
\centering
\footnotesize
\caption{Targeted attack at $K=5$ and $K=10$ when considering a pool of size 128 with $\lambda \in \{0, 0.3\}$. In column AVG we report the average of the measures across all models.}
\resizebox{\textwidth}{!}{%
    \begin{tabular}{|c|c|c|rr|rr|rr|rr|rr|rr|rr|rr|rr|} 
        \cmidrule[\heavyrulewidth]{4-19}
        \multicolumn{1}{l}{} & \multicolumn{1}{l}{\multirow{3}{*}{}} & \multicolumn{1}{l|}{\multirow{2}{*}{}} & \multicolumn{16}{c|}{\textbf{Models}} & \multicolumn{2}{c}{\multirow{1}{*}{}} \\
        \cline{4-21}
        \multicolumn{1}{l}{} & \multicolumn{1}{l}{} & \multicolumn{1}{l|}{} & \multicolumn{2}{c|}{\gemini} & \multicolumn{2}{c|}{\gmn} & \multicolumn{2}{c|}{\zeek} & \multicolumn{2}{c|}{\binfinder} & \multicolumn{2}{c|}{\safe} & \multicolumn{2}{c|}{\jtrans} & \multicolumn{2}{c|}{\trex} & \multicolumn{2}{c|}{\palmtree} & \multicolumn{2}{c|}{AVG} \\
        \cline{3-21}
        \multicolumn{1}{l}{} & \multicolumn{1}{l|}{} & \bm{$\lambda$} & \textbf{0.0} & \textbf{0.3} & \textbf{0.0} & \textbf{0.3} & \textbf{0.0} & \textbf{0.3} & \textbf{0.0} & \textbf{0.3} & \textbf{0.0} & \textbf{0.3} & \textbf{0.0} & \textbf{0.3} & \textbf{0.0} & \textbf{0.3} & \textbf{0.0} & \textbf{0.3} & \textbf{0.0} & \textbf{0.3} \\
        \toprule
        \toprule
        &  & \wasr & 32.17 & 4.84 & 46.30 & 4.03 & 13.98 & 4.35 & 78.18 & 3.63 & 48.20 & 4.37 & 30.80 & 5.88 & 36.40 & 3.74 & 43.29 & 3.50 & 41.16 & 4.29 \\
        \cline{2-21}
        &  & \clean & 11.50 & 11.50 & 8.10 & 8.10 & 7.96 & 7.96 & 6.81 & 6.81 & 8.20 & 8.20 & 11.30 & 11.30 & 5.90 & 5.90 & 7.49 & 7.49 & 8.41 & 8.41 \\
        &  & \asr & 57.60 & 12.50 & 68.70 & 9.94 & 29.80 & 11.74 & 89.28 & 7.52 & 69.50 & 10.94 & 65.30 & 17.40 & 54.10 & 8.13 & 65.17 & 8.95 & 62.43 & 10.89 \\
        &  & \modi & 168.08 & 4.71 & 210.41 & 7.04 & 90.81 & 5.98 & 154.83 & 2.80 & 127.34 & 4.79 & 80.82 & 5.80 & 114.17 & 5.07 & 191.74 & 5.34 & 142.28 & 5.19 \\
        & \multirow{-4}{*}{\textsf{\textbf{@1}}} & \modn & 14.21 & 0.85 & 18.40 & 0.98 & 9.98 & 1.11 & 33.08 & 0.08 & 15.76 & 0.95 & 12.27 & 1.01 & 11.42 & 0.86 & 18.03 & 0.94 & 16.64 & 0.85 \\
\bottomrule
    \end{tabular}}
\vspace{-0.3cm}
\label{tab:Tar_128}
\end{table*}

\begin{table*}[]
\centering
\footnotesize
\caption{Targeted attack at $K=5$ and $K=10$ when considering a pool of size 512 with $\lambda \in \{0, 0.3\}$. In column AVG we report the average of the measures across all models.}
\resizebox{\textwidth}{!}{%
    \begin{tabular}{|c|c|c|rr|rr|rr|rr|rr|rr|rr|rr|rr|} 
        \cmidrule[\heavyrulewidth]{4-19}
        \multicolumn{1}{l}{} & \multicolumn{1}{l}{\multirow{3}{*}{}} & \multicolumn{1}{l|}{\multirow{2}{*}{}} & \multicolumn{16}{c|}{\textbf{Models}} & \multicolumn{2}{c}{\multirow{1}{*}{}} \\
        \cline{4-21}
        \multicolumn{1}{l}{} & \multicolumn{1}{l}{} & \multicolumn{1}{l|}{} & \multicolumn{2}{c|}{\gemini} & \multicolumn{2}{c|}{\gmn} & \multicolumn{2}{c|}{\zeek} & \multicolumn{2}{c|}{\binfinder} & \multicolumn{2}{c|}{\safe} & \multicolumn{2}{c|}{\jtrans} & \multicolumn{2}{c|}{\trex} & \multicolumn{2}{c|}{\palmtree} & \multicolumn{2}{c|}{AVG} \\
        \cline{3-21}
        \multicolumn{1}{l}{} & \multicolumn{1}{l|}{} & \bm{$\lambda$} & \textbf{0.0} & \textbf{0.3} & \textbf{0.0} & \textbf{0.3} & \textbf{0.0} & \textbf{0.3} & \textbf{0.0} & \textbf{0.3} & \textbf{0.0} & \textbf{0.3} & \textbf{0.0} & \textbf{0.3} & \textbf{0.0} & \textbf{0.3} & \textbf{0.0} & \textbf{0.3} & \textbf{0.0} & \textbf{0.3} \\
        \toprule
        \toprule
        &  & \wasr & 16.93 & 1.16 & 26.88 & 1.03 & 5.15 & 1.11 & 58.27 & 0.98 & 27.95 & 0.93 & 15.82 & 1.17 & 20.52 & 0.98 & 18.09 & 1.01 & 23.7 & 1.05 \\
        \cline{2-21}
        &  & \clean & 3.80 & 3.80 & 1.90 & 1.90 & 2.24 & 2.24 & 1.60 & 1.60 & 1.90 & 1.90 & 2.40 & 2.40 & 1.70 & 1.70 & 1.80 & 1.80 & 2.17 & 2.17 \\
        &  & \asr & 34.20 & 3.63 & 49.0 & 2.70 & 12.24 & 3.71 & 71.94 & 2.10 & 49.50 & 2.71 & 38.30 & 3.98 & 34.50 & 2.51 & 34.37 & 2.92 & 40.51 & 3.03 \\
        &  & \modi & 168.49 & 3.19 & 206.50 & 5.81 & 110.12 & 5.22 & 155.03 & 7.86 & 126.75 & 3.11 & 89.29 & 6.32 & 113.79 & 5.48 & 176.89 & 6.93 & 143.36 & 5.49 \\
        & \multirow{-4}{*}{\textsf{\textbf{@1}}} & \modn & 15.67 & 0.44 & 17.87 & 1.04 & 10.58 & 1.0 & 33.09 & 0.10 & 15.64 & 0.59 & 13.28 & 0.90 & 11.32 & 1.20 & 18.37 & 1.17 & 16.98 & 0.80 \\
\bottomrule
    \end{tabular}}

\vspace{-0.3cm}
\label{tab:Tar_512}
\end{table*}

\begin{table*}[]
\centering
\footnotesize
\caption{Targeted attack at $K=5$ and $K=10$ when considering a pool of size 1000 with $\lambda \in \{0, 0.3\}$. In column AVG we report the average of the measures across all models.}
\resizebox{\textwidth}{!}{%
    \begin{tabular}{|c|c|c|rr|rr|rr|rr|rr|rr|rr|rr|rr|} 
        \cmidrule[\heavyrulewidth]{4-19}
        \multicolumn{1}{l}{} & \multicolumn{1}{l}{\multirow{3}{*}{}} & \multicolumn{1}{l|}{\multirow{2}{*}{}} & \multicolumn{16}{c|}{\textbf{Models}} & \multicolumn{2}{c}{\multirow{1}{*}{}} \\
        \cline{4-21}
        \multicolumn{1}{l}{} & \multicolumn{1}{l}{} & \multicolumn{1}{l|}{} & \multicolumn{2}{c|}{\gemini} & \multicolumn{2}{c|}{\gmn} & \multicolumn{2}{c|}{\zeek} & \multicolumn{2}{c|}{\binfinder} & \multicolumn{2}{c|}{\safe} & \multicolumn{2}{c|}{\jtrans} & \multicolumn{2}{c|}{\trex} & \multicolumn{2}{c|}{\palmtree} & \multicolumn{2}{c|}{AVG} \\
        \cline{3-21}
        \multicolumn{1}{l}{} & \multicolumn{1}{l|}{} & \bm{$\lambda$} & \textbf{0.0} & \textbf{0.3} & \textbf{0.0} & \textbf{0.3} & \textbf{0.0} & \textbf{0.3} & \textbf{0.0} & \textbf{0.3} & \textbf{0.0} & \textbf{0.3} & \textbf{0.0} & \textbf{0.3} & \textbf{0.0} & \textbf{0.3} & \textbf{0.0} & \textbf{0.3} & \textbf{0.0} & \textbf{0.3} \\
        \toprule
        \toprule
        &  & \wasr & 11.35 & 0.55 & 19.80 & 0.65 & 0.0 & 0.0 & 49.55 & 0.45 & 19.60 & 0.43 & 11.0 & 0.40 & 15.17 & 0.55 & 13.83 & 0.55 & 17.54 & 0.45 \\
        \cline{2-21}
        & \multirow{4}{*}{} & \clean & 1.90 & 1.90 & 0.90 & 0.90 & 0.0 & 0.0 & 0.90 & 0.90 & 1.0 & 1.0 & 0.90 & 0.90 & 1.20 & 1.20 & 0.70 & 0.60 & 0.94 & 0.94 \\
        &  & \asr & 24.30 & 1.81 & 38.60 & 1.50 & 0.0 & 0.0 & 63.63 & 1.10 & 38.50 & 1.31 & 27.20 & 1.50 & 27.60 & 1.51 & 26.85 & 1.51 & 30.84 & 1.28 \\
        &  & \modi & 167.52 & 1.44 & 205.78 & 4.67 & - & - & 154.96 & 7.36 & 129.22 & 4.23 & 94.96 & 3.67 & 115.66 & 3.6 & 178.75 & 7.67 & 149.55 & 4.66 \\
        & \multirow{-4}{*}{\textsf{\textbf{@1}}} & \modn & 15.69 & 0.33 & 17.11 & 1.07 & - & - & 33.31 & 0.0 & 15.98 & 0.77 & 13.80 & 0.53 & 11.11 & 0.67 & 19.0 & 0.93 & 18.0 & 0.61 \\
\bottomrule
    \end{tabular}}
\label{tab:Tar_1000}
\vspace{-0.3cm}
\end{table*}

\begin{table*}[]
\centering
\footnotesize
\caption{Targeted attack at $K=5$ and $K=10$ when considering $\lambda=0.01$ and $|\mPool| \in \{32, 128, 512, 1000\}$. In column AVG we report the average of the measures across all models.}
\resizebox{\textwidth}{!}{%
    \begin{tabular}{|l|l|c|rrrr|rrrr|rrrr|rrrr|}
    \cmidrule[\heavyrulewidth]{4-15}
    \multicolumn{1}{l}{} & \multicolumn{1}{l}{} & \multicolumn{1}{l|}{} & \multicolumn{12}{c|}{\textbf{Models}} & \multicolumn{1}{c}{} & \multicolumn{1}{c}{} & \multicolumn{1}{c}{} & \multicolumn{1}{c}{} \\
    \cline{4-19}
    \multicolumn{1}{l}{} & \multicolumn{1}{l}{} & \multicolumn{1}{l|}{} & \multicolumn{4}{c|}{\gemini} & \multicolumn{4}{c|}{\safe} & \multicolumn{4}{c|}{\jtrans} & \multicolumn{4}{c|}{AVG} \\
    \cline{3-19}
    \multicolumn{1}{l}{} & \multicolumn{1}{l|}{} & \textbf{Pool Size} & 32 & 128 & 512 & 1000 & 32 & 128 & 512 & 1000 & 32 & 128 & 512 & 1000 & 32 & 128 & 512 & 1000 \\
    \toprule
    \toprule
    &  & \wasr & 29.49 & 12.52 & 4.56 & 2.48 & 54.23 & 28.25 & 13.63 & 8.38 & 32.17 & 10.80 & 3.33 & 1.80 & 38.63 & 17.19 & 7.17 & 4.22 \\
    \cline{2-19}
    & \multirow{4}{*}{} & \clean & 30.26 & 11.90 & 3.97 & 2.08 & 27.0 & 8.30 & 1.90 & 1.0 & 42.60 & 11.10 & 2.40 & 0.90 & 33.29 & 10.43 & 2.76 & 1.33 \\
    &  & \asr & 53.08 & 26.69 & 11.41 & 5.95 & 73.30 & 45.80 & 26.40 & 18.50 & 70.10 & 29.70 & 9.70 & 4.90 & 65.49 & 34.06 & 15.84 & 9.78 \\
    &  & \modi & 11.36 & 12.16 & 13.55 & 13.83 & 24.53 & 28.12 & 30.52 & 30.85 & 7.34 & 8.42 & 8.80 & 7.88 & 14.41 & 16.23 & 17.62 & 17.52 \\
    & \multirow{-4}{*}{\textsf{\textbf{@1}}} & \modn & 3.67 & 4.13 & 4.61 & 4.80 & 4.86 & 5.28 & 5.62 & 5.65 & 1.84 & 2.11 & 2.21 & 1.88 & 3.46 & 3.84 & 4.15 & 4.11 \\
\bottomrule
    \end{tabular}}

\vspace{-0.3cm}
\label{tab:Tar_001}
\end{table*}

% \section{}\label{sec:app_B}

% To avoid potential misuse of our pipeline, particularly in malware analysis, the code will not be open-sourced. However, it will be accessible to interested members of the scientific community, subject to appropriate checks.